\documentclass[a4paper,11pt]{article}
\pdfoutput=1
\usepackage{jcappub} 
\usepackage{footnote}	
\usepackage{xcolor}
\usepackage{bigints}
\usepackage{multirow}
\usepackage{appendix}
\usepackage[normalem]{ulem}
\usepackage{bm,relsize}        
\usepackage{amsfonts,amsmath,amssymb,amsthm,mathtools}
\def\DM		{{\mathsmaller{\text{DM}}}}

\def\NFW		{{\mathsmaller{\text{NFW}}}}

\graphicspath{{figs/}} 

\begin{document}

\title{Dark Matter spikes around Sgr A* in $\gamma$-rays
}

\author[a,b]{Shyam Balaji,}
\author[c]{Divya Sachdeva,}
\author[d,1]{Filippo Sala\note{On leave of absence from LPTHE, CNRS \& Sorbonne Universit\'{e}, Paris, France.}}
\author[b,e,f]{and Joseph Silk}

\affiliation[a]{Laboratoire de Physique Th\'{e}orique et Hautes Energies (LPTHE), \\
	UMR 7589 CNRS \& Sorbonne Universit\'{e}, 4 Place Jussieu, F-75252, Paris, France}
	\affiliation[b]{Institut d’Astrophysique de Paris, UMR 7095 CNRS \& Sorbonne Universit\'{e}, 98 bis boulevard Arago, F-75014 Paris, France}
 \affiliation[c]{Laboratoire de Physique de l'\'{E}cole normale sup\'{e}rieure, ENS, Universit\'{e} PSL, CNRS, Sorbonne Universit\'{e}, Universit\'{e} Paris Cit\'{e}, 24 Rue Lhomond, F-75005 Paris, France}
 \affiliation[d]{Dipartimento di Fisica e Astronomia, Università di Bologna
and INFN sezione di Bologna, Via Irnerio 46, I-40126 Bologna, Italy}
\affiliation[e]{Department of Physics and Astronomy, The Johns Hopkins University, 3400 N. Charles	Street, Baltimore, MD 21218, U.S.A.}
\affiliation[f]{Beecroft Institute for Particle Astrophysics and Cosmology, University of Oxford, Keble	Road, Oxford OX1 3RH, U.K.}

\emailAdd{sbalaji@lpthe.jussieu.fr}
\emailAdd{divya.sachdeva@phys.ens.fr}
\emailAdd{f.sala@unibo.it}
\emailAdd{silk@iap.fr}

\abstract{
We use H.E.S.S. $\gamma$-ray observations of Sgr A* to derive novel limits on the Dark Matter (DM) annihilation cross-section. We quantify their dependence on uncertainties i) 
in the DM halo profile, which we vary from peaked to cored, and ii) in the shape of the DM spike around Sgr A*, dynamically heated by the nuclear star cluster. For peaked halo profiles and depending on the heating of the spike, our limits are the strongest existing ones for DM masses above a few TeV. Our study contributes to assessing the influence of the advancements in our knowledge of the Milky Way on determining the properties of DM particles.
}

\maketitle

\section{Introduction}

The identification of the constituents of dark matter (DM) is one of the most important open questions in fundamental physics, and sits at the intersection of astrophysics, cosmology and particle physics.
From the particle physics point of view, the  hypothesis that DM is made of 
weakly interacting particles 
with a mass $m_\chi$ at or above the TeV scale is well motivated.
Colliders have not only excluded many DM candidates with a mass below a few hundreds of GeV~\cite{Boveia:2018yeb}, but they have also pushed the scale of many new physics sectors to larger energies~\cite{Hinzmann:2022okt}, and DM could naturally exist in this regime as part of these sectors.

Cosmology independently motivates $m_\chi \gtrsim $~TeV via the DM production mechanism of thermal freeze-out, which sets the observed DM abundance when its interactions with the early-universe bath 
fall out of equilibrium.
This production mechanism implies that DM annihilation should generate signals at telescopes, that  already have excluded the simplest realisations of this picture for $m_\chi < \mathcal{O}(100)$~GeV, see e.g. Ref.~\cite{Slatyer:2021qgc}. Larger masses are within reach of current and future telescopes and hence are of particular experimental interest.

Astrophysics provides information about the DM mass density $\rho_\DM$, and thus gives crucial input for telescope searches of DM annihilation signals, as these are proportional to $\rho_\DM^2$. 
It is therefore promising to investigate the Galactic Center (GC) region of the Milky Way (MW), because $\rho_\DM$ is expected to increase in this direction due to the strong gravitational potential.
In addition the supermassive black hole (SMBH) at the GC, Sgr A*, has had the time to possibly accrete DM in its proximity into a `spike' with density $\rho_\text{spike} \gg \rho_\DM$, that could then source a large annihilation signal~\cite{Gondolo:1999ef}.

\medskip

Over the last decade, we have successfully detected multi-TeV $\gamma$-rays emanating from Sgr A* for the first time using telescopes such as H.E.S.S.~\cite{HESS:2016pst}, VERITAS~\cite{Adams:2021kwm}, and MAGIC~\cite{MAGIC:2020kxa}. With advancements in detectors, such as LHAASO~\cite{LHAASO:2019qtb} and the upcoming CTA~\cite{CTAConsortium:2019emb,CTA:2020qlo}, we anticipate further advancements in this area. It is, therefore, critical to analyse the aforementioned data to determine, in combination with other existing DM searches, novel insights about both DM annihilation and the dark matter mass distribution.

\medskip

The knowledge of the DM density profile in the spike is affected by severe uncertainties.
Some come from the poor knowledge of $\rho_\DM$ in the inner kpc of the Milky Way (MW), which is a necessary ingredient to compute $\rho_\text{spike}$. 
Indeed baryons dominate the MW dynamics in that region, making it hard to establish $\rho_\DM$ from observations or from simulations.
Recent progress in both directions (see e.g.~\cite{Pato:2015dua,2020MNRAS.494.4291C} on the observational side and~\cite{DiCintio:2014xia,Sameie:2021ang} for simulations) has however enabled us to restrict the realm of possibilities with respect to that of only a decade ago.
Irrespectively of future astrophysical progress on $\rho_\DM$, an annihilation signal could be observed by telescopes, from the region of Sgr A* and/or from much more extended regions around the GC.
It is therefore desirable to know how the strength of the signals, from a DM spike around Sgr A* versus {{ from}  more extended halo regions, correlate, as a function of some meaningful parametrisation of astrophysical uncertainties.

Besides the uncertainties caused by the DM halo in close proximity to the GC that were discussed earlier, the density profile of the DM spike also introduces uncertainties regarding the impact of baryonic matter, such as stars, on the DM in the vicinity of Sgr A*. However, these uncertainties are being reduced as our comprehension of the baryonic matter in the region surrounding Sgr A* continues to improve.
Dramatic observational progress has recently been achieved, for example, on the orbits of stars closest to Sgr A*~\cite{GRAVITY:2021xju,Heissel:2021pcw} and on the nuclear star cluster profile~\cite{2019Habibi,pechetti2020luminosity,2020Gallego}.
Many of these observational results were unknown at the time of the first studies that aimed to determine $\rho_\text{spike}$, such as~\cite{Ullio:2001fb,Gnedin:2003rj,Bertone:2005hw}.
Therefore, at present, we know which assumptions in these studies do not stand observational scrutiny, and moreover novel determinations of $\rho_\text{spike}$ have { subsequently}  been possible, see e.g. Ref.~\cite{Shapiro:2022prq}.

\medskip

Given recent and upcoming advancements on the observation of multi-TeV $\gamma$-rays from Sgr A*, as well as in understanding the density of DM in the GC and the spike, it is a timely moment to evaluate their impact on our knowledge of DM annihilation properties.
This paper performs a step in that direction, by looking for DM annihilation signals from the region of Sgr A* in public H.E.S.S. data~\cite{HESS:2016pst} on high-energy $\gamma$-rays, and interpreting the results for various astrophysically-motivated spike DM profiles.

\section{Dark matter spikes}
\label{sec:DMspikes}

\subsection{Dark matter halo profile down to Sgr A*}
The SMBH Sgr A* exerts a gravitational influence that, over the entire history of the MW, can dramatically increase the density profile of DM in close proximity compared to its halo component without an SMBH.
This denser-than-halo DM profile, close to Sgr A*, is called a `spike'. 
The first proposal to look at the spike with telescopes in search for enhanced DM annihilation signals, by Gondolo and Silk~\cite{Gondolo:1999ef}, dates back to the nineties.
Since then, the knowledge of the physics that determines the spike has substantially improved, and we employ it in this section to define and motivate benchmarks for the DM spike profile. They will be useful to disentangle the impact of the various astrophysical uncertainties on the strength of DM annihilation signals.

We employ the following parametrisation of the DM mass density of the MW
\begin{align}
\label{eq:spikedensity}
    \rho(r)= \left\{
        \begin{array}{ll}
            0 & \quad r<2 R_S \\
            \displaystyle\rho_\textrm{sat}\Big(\frac{r}{R_{\rm sat}}\Big)^{-0.5} & \quad 2 R_S \leq r < R_\textrm{sat} \\
            \displaystyle\rho_\textrm{spike}(r) & \quad R_\textrm{sat} \leq r < R_\textrm{sp} \\
            \rho_\textrm{halo}(r) & \quad r \geq R_\textrm{sp}
        \end{array} ,
    \right.
\end{align}
where
\begin{itemize}
    \item[$\diamond$] $r$ is the distance from Sgr A*, which we assume to be at the exact center of the MW (we discuss this assumption in Sec~\ref{sec:spikeDM});
    \item[$\diamond$]
    $\rho_\textrm{halo}(r)$ is the halo DM mass density. We discuss it in Sec.~\ref{sec:haloDM} and give it in Eqs.~(\ref{eq:NFW}) and~(\ref{eq:coredspikedensity});
    \item[$\diamond$] $R_\textrm{sp}$ and $\rho_\textrm{spike}$ are, respectively, the radial extension and the mass density profile of the DM spike. They are discussed in Sec.~\ref{sec:spikeDM} and given in Eqs.~(\ref{eq:Rspike}) and~(\ref{eq:DMspike});
    \item[$\diamond$] $R_\textrm{sat}$ and $\rho_\textrm{sat}$ are, respectively, the saturation radius and density of the spike due to DM annihilation. They are discussed in Sec.~\ref{sec:satDM} and given in Eqs~(\ref{eq:Rsat}) and~(\ref{eq:rhosat});
    \item[$\diamond$]
    $R_S=2GM_\textrm{BH}/c^2=2.95\,(M_\textrm{BH}/M_\odot)\textrm{\,km}$ is the Schwarzschild radius of the BH.
    \end{itemize}
Our benchmarks for the halo plus spike DM profiles, that we will discuss next, are displayed in Figure~\ref{fig:spikeprofiles}.

\begin{figure}[t]
    \centering
    \includegraphics[width=7cm]{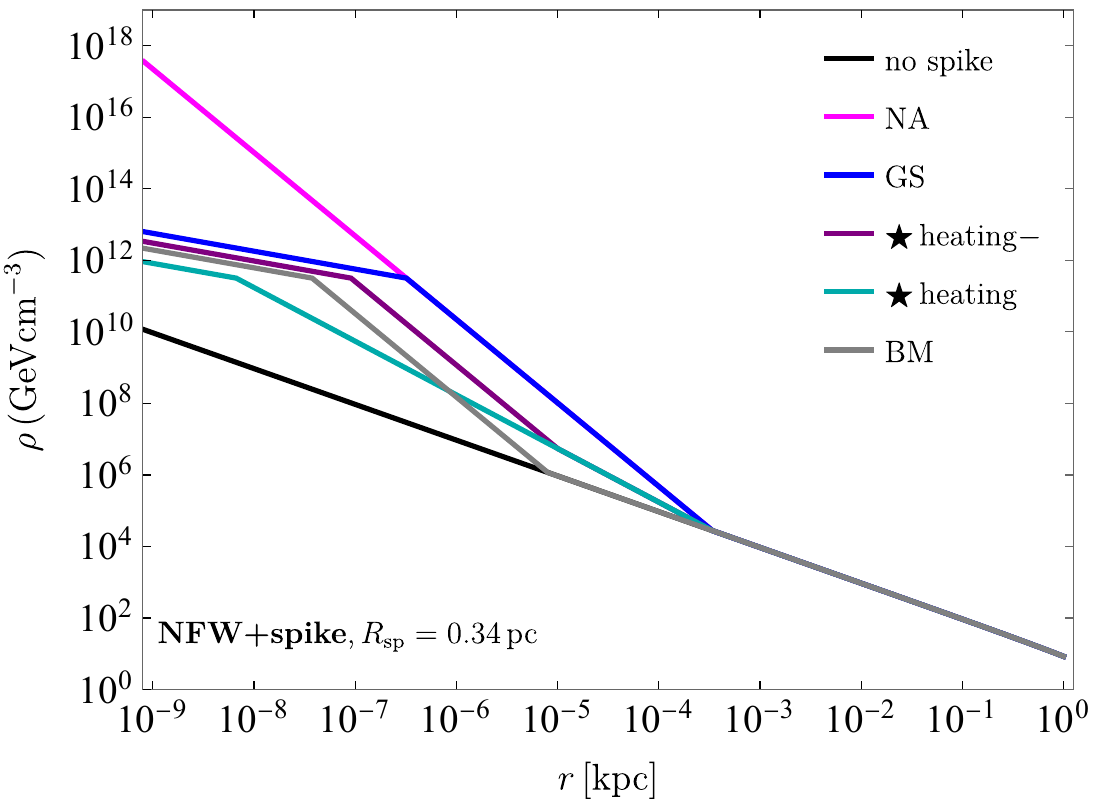}
    \includegraphics[width=7cm]{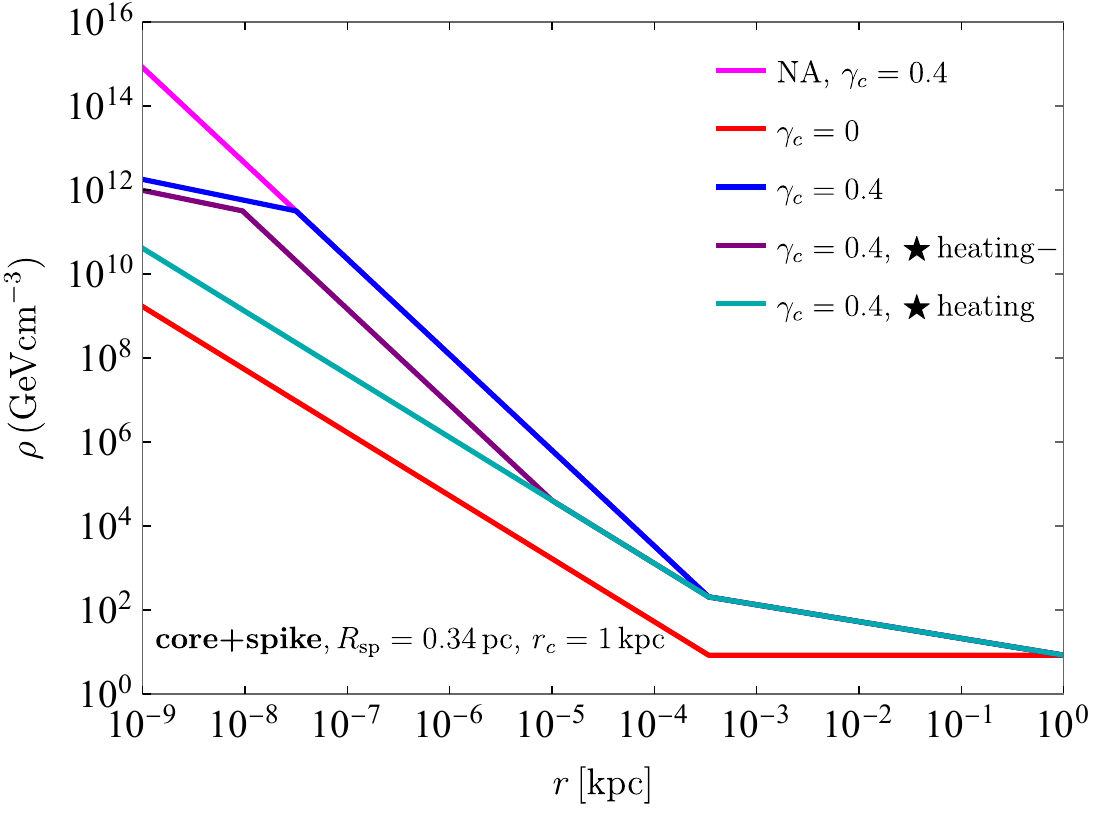}
    \caption{
    Dark Matter density profiles in the Milky Way for a NFW halo with a spike (left-hand panel) and cored halo with a spike (right-hand panel), as a function of the distance $r$ from Sgr A*.
    In both panels, the spike radius $R_\textrm{sp}$ is  0.34~pc.
    Lines in different colors correspond to the different benchmarks discussed in the main text.
    For the NFW panel, they are: non-annihilating (NA) (magenta), Gondolo-Silk (GS) (blue), less stellar heating (purple), stellar heating (cyan) and Bertone-Merritt (BM) (gray).
    For the cored panel, they are: $\gamma_c=0.4$ with NA (magenta), GS (blue), less stellar heating (purple), stellar heating (cyan), and $\gamma_c=0$ (red) which is the same for all the above spike benchmarks. We set $m_\chi=1$~TeV and $\langle \sigma v \rangle=10^{-26}\,\textrm{cm}^{-3}{\textrm {s}}^{-1}$ for each case with dark matter annihilation. All cored profiles have a core radius $r_c=1$ kpc.
    }
    \label{fig:spikeprofiles}
\end{figure}

\subsection{Dark matter in the halo}
\label{sec:haloDM}

Because the slope of a spike is strongly dependent on the halo profile close to the BH, it is critical to account for the uncertainties in the DM halo distribution at small Galactocentric radii.
While DM-only N-body simulations suggest that the DM density in the halo peaks in the galactic center~\cite{Navarro:2008kc,Diemand:2008in}, 
the existence of baryons may, via a number of ways, drastically alter the DM distribution. These include feedback \cite{Mashchenko:2007jp,Maccio:2011ryn,Governato:2012fa}, which tends to flatten the DM density profile, and adiabatic contraction~\cite{Gnedin:2004cx,Gnedin:2011uj,Pato:2015dua}, which increases the DM density around the GC.
Given that the role of baryonic effects is still a topic of debate, in what follows we will consider both peaked and cored profiles.\footnote{
The DM halo profile can also be flattened by altering the DM particle characteristics, such as by introducing self-interactions \cite{Rocha:2012jg,Kaplinghat:2013xca}, a possibility that we will not include in the rest of this paper.}


As a peaked profile, we employ that of  NFW~\cite{Navarro:1995iw}, given as
\begin{equation}
\label{eq:NFW}
    \rho_\textrm{halo}^\NFW(r)
    = \rho_s\left(\frac{r}{r_s}\right)^{-1} \left(1+\frac{r}{r_s}\right)^{-2} \,,
\end{equation}
with $r_s=18.6\,$kpc and $\rho_s\,=\,\displaystyle\rho_\odot(R_\odot/r_s) (1+R_\odot/r_s)^{2}$, where $R_\odot\,=\,8.2\,\,{\rm kpc}$ is the sun position and $\rho_\odot=0.383\, \textrm{GeV}/\textrm{cm}^3$ is the local DM density~\cite{mcmillan2016mass}.

Coming to cored density profiles, we employ the following parameterisation:
\begin{align}
\label{eq:coredspikedensity}
    \rho_\textrm{halo}^{\rm cored}(r,\gamma_c)= \left\{
        \begin{array}{ll}
            \rho_\textrm{halo}^\NFW(r_c)\,
            \left(\dfrac{r}{r_c}\right)^{-\gamma_c}
            &\quad  r < r_\textrm{c} \\
           \rho_\textrm{halo}^\NFW(r) & \quad r \geq r_\textrm{c}
        \end{array} ,
    \right.
\end{align}
where $0 \leq\gamma_c\,<\,1$ and $r_c$ is the core radius. For Milky Way-sized galaxies, simulations allow for core radii of the order of a kpc~\cite{Chan:2015tna,Sameie:2021ang}, with previous simulations finding even larger cores~\cite{Mollitor:2014ara}. In this work, we assume $r_c\,=\,1\,{\rm kpc}$ for definiteness. Since baryons largely dominate the gravitational potential at those scales, in Eq.~(\ref{eq:coredspikedensity}) we can safely use the same NFW parameters~\cite{mcmillan2016mass} that we used for Eq.~(\ref{eq:NFW}).
We consider two possibilities for $\gamma_c$:
\begin{itemize}
    \item $\gamma_c\,=\,0.4$, as a representative value of the peak softening found below 1~kpc by the recent FIRE simulation of baryons plus cold DM~\cite{Sameie:2021ang};
    \item $\gamma_c\,=\,0$, to be conservative.
\end{itemize}

Our choice of $\rho_\odot = 0.383\, \textrm{GeV}/\textrm{cm}^3$ is conservative, because it is smaller than that favoured both by local GAIA measurements (e.g.~\cite{Evans:2018bqy} finds $\rho_\odot = 0.55 \pm 0.17$~GeV/cm$^3$, \cite{Buch:2018qdr} finds even larger values), and by global determinations (e.g.~\cite{Pato:2015dua} finds $\rho_\odot \simeq 0.42 \pm 0.03$~GeV/cm$^3$ by using thousands of MW measurements and 70 possibilities for the baryonic modelling). 
Mild variations in~$\rho_\odot$ will turn out to have a stronger impact on our limits than in usual indirect detection searches, so we will present limits also for the case $\rho_\odot = 0.55$~GeV/cm$^3$.


\subsection{Dark matter in the spike}
\label{sec:spikeDM}

Numerical studies~\cite{Gnedin:2003rj,Merritt:2003qc} suggest that, fairly independently of its density profile, the spike begins to grow around the radius $R_{\rm sp} \simeq 0.2\,R_h$, where $R_h = GM_{\rm BH}/v_o^2$ is the gravitational influence radius of the BH and $v_o$ is the velocity dispersion of the stars in the inner halo. $R_h$ is defined by the condition that the potential energy due to the BH is equal to the typical kinetic energy of a DM particle in the halo.
    Adopting $M_{\rm BH}\,=\,4.3\times 10^6\,{\rm M}_\odot$ for Sgr A$^*$~\cite{GRAVITY:2021xju} and $v_o\,=\, 105 \pm 20\, {\rm km\, s}^{-1}$~\cite{2009ApJ...698..198G}, we find for the MW
    \begin{equation}
    \label{eq:Rspike}
        R_h\,=\,1.7\,{\rm pc}\,,\qquad R_{\rm sp}\,\simeq\,0.34\,{\rm pc}\,.
    \end{equation}

We employ the following parametrisation for the mass density profile of the DM spike
\begin{equation}
\label{eq:DMspike}
\rho_\textrm{spike}(r)\, = \, \rho_\textrm{halo}(R_\textrm{sp})\left(\frac{r}{R_\textrm{sp}}\right)^{-\gamma_\textrm{sp}(r,\gamma)}\,,
\end{equation}
and proceed to define and discuss benchmarks and their limitations.
We anticipate that, for values of $R_{\rm sp}$ in the ballpark of Eq.~(\ref{eq:Rspike}), neither radio interferometry nor astronomical observations of the star S2 orbiting Sgr A$^*$ are (yet) sensitive enough to gravitationally detect the presence of the DM spikes we will consider in our study, see e.g.~\cite{GRAVITY:2021xju,Nampalliwar:2021tyz,Shen:2023kkm}.

\paragraph{Gondolo-Silk (GS).}
Under the assumption of adiabatic growth of a peaked DM halo density around  a central SMBH,
the value of $\gamma_\textrm{sp}$ is predicted as a function of the halo slope as~\cite{Gondolo:1999ef,Ullio:2001fb}
\begin{equation}
\label{eq:spikeslope}
    \gamma_\textrm{sp}(\gamma>0)=\displaystyle\frac{9-2\gamma}{4-\gamma}\,,
\end{equation}
so that $2.4 \geq \gamma_\textrm{sp} > 2.25$ for $1.5 \geq \gamma > 0$.
Eq.~(\ref{eq:spikeslope}) assumes the DM halo profile is `peaked' (i.e. $\gamma > 0$) close to the central BH.
For cored (i.e. $\gamma = 0$) halo profiles  close to the central BH, $\gamma_\textrm{sp}$ takes a value of either 2 or 1.5, depending on other parameters of the DM halo distribution, see e.g.~\cite{Ullio:2001fb}. To be conservative, we assume
\begin{equation}
\label{eq:spikecore}
\gamma_\textrm{sp}(\gamma = 0) = 1.5\,.
\end{equation}
This value is very conservative for one more reason, namely simulations~\cite{Chan:2015tna}  show that cores may form close to $z\,\sim\,0$, while the steepness of the spike depends on the shape of the halo at earlier times too, when it was not cored yet. So one may well expect $\gamma_\textrm{sp} > 1.5$.

\paragraph{Stellar heating ($\bigstar$ heating).}
The picture above may well be altered by the baryons (from stars etc.) surrounding Sgr A*.
In fact, gravitational interactions between DM and the baryons dampen the spike.
For instance, the DM spike may be significantly softened due to gravitational `heating' by the nuclear star cluster, i.e. the stars within the inner few pc of the MW.
This results in an equilibrium spike solution as low as~\cite{Gnedin:2003rj,Bertone:2005hw,Merritt:2006mt,Shapiro:2022prq} 
\begin{equation}
\gamma_{\rm sp}^{\rm heated}\,=\,1.5
\end{equation}
($\gamma_{\rm sp}\,=\,1.5$ and $\gamma_{\rm sp}\,=\,1.8$ after 10 Gyr and 20 Gyr~\cite{Bertone:2005hw}). 
We call this benchmark `stellar heating'.

As far as we know, the studies carried out on the dampening of spikes have all been in the context of NFW profiles. 
For cored profiles, we also choose to work with $\gamma_{\rm sp}^{\rm heated}\,=\,1.5$, because we start anyway from a slope of 1.5 for the unheated cored profile Eq.~(\ref{eq:spikecore}) and the nuclear star cluster has a density slope of $\gamma_{\rm NSC} \simeq 1.4$~\cite{2020Gallego}, so we do not expect it to heat the DM to flatter slopes.

%
\paragraph{Bertone-Merritt (BM).}
The non-equilibrated spikes undergoing a stellar heating process and the spike signal reduction can, alternatively, be parameterised as follows~\cite{Ahn:2007ty}:
\begin{equation}
\label{depletedspike}
R_{\rm sp}(t)\,=\,R_{\rm sp}(0)e^{-\tau/2(\gamma_{\rm sp}-\gamma)} 
\end{equation}
where $\gamma_{\rm sp}=\gamma_{\rm sp}(\gamma)$ of Eqs~(\ref{eq:spikeslope})-(\ref{eq:spikecore}) and $\tau\,=\,10$ is the time since the spike formed in units of the heating time $T_{\rm heat}\,=\,1.25\times\,10^9\,{\rm yrs}$~\cite{Ahn:2007ty,Merritt:2003qk}.
This approach parametrises the effect of DM heating with a decrease in $R_{\rm sp}$, rather than a decrease in $\gamma_{\rm sp}$ with $R_{\rm sp}$ remaining unchanged.
However, as discussed in Ref.~\cite{Sandick:2016zeg} and shown  in the next section, both approaches produce indistinguishable results for the spike signal from a NFW profile.
We call this benchmark `Bertone-Merritt (BM)'.

\paragraph{Less stellar heating ($\bigstar$ heating$-$).}
We also consider an alternative spike softening scenario, where the DM spike follows  the Gondolo-Silk relation for $\gamma_\textrm{sp}$ (given in Eqn.~\ref{eq:spikeslope}) within the milliparsec region and varies as $r^{-\gamma_{\rm sp}^{\rm heated}}$ for $r\,>\,0.01$ pc. Our reasoning is as follows: based on the stellar density inferred in the inner $0.04\,{\rm pc} \,\times\, 0.04\,{\rm pc}$ region in Refs~\cite{2019Habibi} and \cite{2020Gallego}, there are only a few older/brighter giants in this region, and the mean separation between the nuclear  cluster stars can be estimated to be of $\mathcal{O}(10^{-2})$ pc~\footnote{Note that the measurement of stellar density and consequently the mean separation of NSC's in the inner region is subject to the current angular resolution. These numbers could change in future studies that employ imaging and spectroscopy with higher angular resolution.}.  Thus, naively, there should be no scattering between DM and stars in $r\,<\,0.01$ pc. Hence, the DM spike should theoretically not be softened within this region. 
We call this benchmark `less stellar heating'.

One may question the apparent paucity of stars in the inner $0.01$~pc, since it seems highly plausible that both DM and stars  follow a spiky profile. To counter that, we can consider the scenario where the BH grows (say by gas accretion) mostly before the nuclear star cluster forms in the spike region. In that case, the DM and stellar profiles are decoupled. Additionally, the formation of nuclear star clusters may be due to the merger of globular clusters, as suggested in Ref.~\cite{2015Antonini}. This may in turn lead to different profiles for the DM and stellar distributions. Therefore, stars and DM could be essentially decoupled and the absence of older/ brighter giants near GC can prevent softening of DM spike in the 
milliparsec region.

\paragraph{Limitations of our benchmarks.}

The benchmarks defined above encompass many of the uncertainties affecting the determination of spike profiles.
We list here other uncertainties that we have not accounted for, together with their expected impact.

We have not defined benchmarks with halo profiles steeper than NFW, despite {the case that} some simulations do produce a DM profile steeper than NFW, when correlated with Gaia data of the baryonic content of the MW~\cite{2020MNRAS.494.4291C}. If we did include such benchmarks, they would of course yield to stronger limits.

Concerning the spike radius Eq.~(\ref{eq:Rspike}) note that, by using the velocity dispersion of the stars to determine it, we have implicitly assumed that DM and the baryons are at kinetic equilibrium. This holds in all benchmark we choose, but could perhaps be too conservative for GS, where we neglect DM heating by stars and so DM could have a smaller velocity dispersion, leading to a larger spike radius. Since this observation carries some further uncertainties, given that the GS benchmarks lead to the strongest constraints anyway, for simplicity (and in the same spirit of the previous limitation) we decided to use the same spike radius for all benchmarks.

If the GC underwent mergers with other galaxies, then the shape of the spike would be affected~\cite{Gnedin:2003rj,Ullio:2001fb}.
However the MW is unlikely to have undergone any major merger in its recent past (see e.g.~\cite{Bertone:2005hw,Lacroix:2018zmg}) and 
GAIA data actually suggests that the MW underwent its last  major merger, the so-called GAIA Sausage/Enceladus event, several billion years ago \cite{ciucua2022chasing}.
The spike therefore has had time to regenerate, plus it could be affected, in any direction, by the thick disk resulting from the merger.
 We do not account for major merger effects in our modelling of $\rho_\textrm{spike}$.

Another assumption common to all our benchmarks is that Sgr A*, throughout most of its formation history, sits within tens of pc of the center ($r=0$) of $\rho_\textrm{halo}$.
While this assumption is in principle not guaranteed (see e.g.~\cite{Ullio:2001fb}), the fact that Sgr A* sits at the center of the baryon distribution of the MW now receives indirect support by the measured slope of the density of the nuclear star cluster, see e.g.~\cite{2020Gallego}.
Since in these areas, the gravitational dynamics is by far dominated by baryons, not by DM, we do not find it unreasonable to assume that Sgr A* has sit at the center of $\rho_\textrm{halo}$ for most of its lifetime. We therefore do not account for departures from this assumption in our modelling.
If one were to do this, then we would expect spikes built on top of peaked profiles  to be softened.
Concerning cored halo profiles with $\gamma_c=0$, they are instead somehow immune to this assumption: even if Sgr A* moved from the center it would induce the same $\rho_\textrm{spike}$, provided it stays within the core and it moved sufficiently slowly.


\subsection{Saturation due to dark matter annihilations}
\label{sec:satDM}
    $R_\textrm{sat}$ is the saturation radius of the spike due to DM annihilations.
    We discuss this separately from the rest of the spike because it depends mostly on DM physics and very indirectly on baryonic effects.
    For DM with mass $m_\chi$ and annihilating with cross-section $\langle\sigma v\rangle$, the density growth in the innermost region of the DM spike is depleted because DM annihilates efficiently on account of very high DM density. This softening of the spike happens at $r\,<\,R_\textrm{sat}$, where $R_\textrm{sat}$ is the radius where $\rho_\textrm{sat}=\displaystyle\rho_\textrm{spike}$
    \begin{equation}
    \label{eq:rhosat}
    \rho_{\rm sat} = \frac{m_\chi}{\langle\sigma v\rangle t_{\rm BH}}
    \simeq 3.17\times\,10^{11}\,{\rm GeV}\,{\rm cm}^{-3}\frac{m_\chi}{10~{\rm TeV}} \frac{10^{-25} \textrm{cm}^3/\text{s}}{\langle\sigma v\rangle}\frac{10^{10}\textrm{yr}}{t_{\rm BH}}\,,
\end{equation}
where we use $t_\textrm{BH}=10^{10}$yr as the age of the BH. $R_\textrm{sat}$ is defined by
\begin{equation}
\label{eq:Rsat}
\rho_\textrm{sat}
=\displaystyle\rho_\textrm{halo}(R_\textrm{sp})\left(\frac{R_\textrm{sat}}{R_\textrm{sp}}\right)^{-\gamma_\textrm{sp}(r,\gamma)},
\end{equation}
which ensures continuity of the DM distribution. Since the trajectories of DM constituents close to Sgr A* are not circular, within $r\,<\,R_{\rm sat}$ the DM distribution forms a weak cusp, characterised by $r^{-0.5}$ for $s$-wave annihilation~\cite{Vasiliev:2007vh,Shapiro:2016ypb}, rather than a plateau.

\section{$\gamma$-ray flux from dark matter annihilation}
\label{sec:flux}
WIMP DM particles are expected to annihilate with each other into SM particles, that eventually produce primary (i.e. at the particle physics level) $\gamma$-rays.
The photon energy-dependent prompt $\gamma$-ray flux, from annihilations of self-conjugate DM, can be calculated as
\begin{align}
    \frac{d\Phi}{dE_\gamma}=\frac{\langle \sigma v\rangle}{8\pi m_\chi^2} \frac{dN_\gamma}{dE_\gamma} J
\end{align}
where $m_\chi$ is the rest mass of the DM particle, $\langle \sigma v\rangle$ is the thermally-averaged cross section for the DM annihilation, $\frac{dN_\gamma}{dE_\gamma}$ represents the differential photon energy spectrum for each DM annihilation (calculated with the PPPC4DMID package \cite{Cirelli:2010xx,Ciafaloni:2010ti}), and the $J$ factor  encodes the spatial distribution of the DM density.

The energy spectrum of the photons produced by one DM annihilation can be written as a sum over all possible final states
\begin{align}
    \frac{dN_\gamma}{dE}=\sum_i \mathcal{B}_i \frac{dN_i}{dE_\gamma} 
\end{align}
where $\mathcal{B} = \langle \sigma v\rangle_i/\langle \sigma v\rangle$ is the branching fraction and $i$ denotes each channel, for example $i =b\bar{b}, W^+ W^-$.
The $J$ factor is calculated~as
\begin{align}
\label{eq:jfactor}
    J=\int_{\Delta\Omega} \int_\textrm{l.o.s} \rho^2(\mathbf{r}(\psi,l)(s)) dl d\Omega
\end{align}
where $\psi$ is the angle between the direction of the GC and the line-of-sight, $\mathbf{r}(\psi,l)=\sqrt{r_\odot^2-2lr_\odot \cos\psi+l^2}$, and the line-of-sight distance $l$ is the variable we integrate over.
We also  require the maximum $l_\textrm{max}=\sqrt{R^2-r_\odot^2\sin^2\psi}+r_\odot \cos\psi$  as the upper boundary for the integral over $l$ in Eq.~\eqref{eq:jfactor}.
The radius $R=200$~kpc is the halo's virial radius, the $J$ factor is largely insensitive to its exact value.

For DM spikes, the $J$ factor is impacted by both $m_\chi$ and $\langle\sigma\,v\rangle$ through $\rho_{\rm sat}$, leading to changes in the derived limits on DM annihilation that are not directly proportional to the $J$ factor at fixed $\langle \sigma v\rangle$. To illustrate this point, we calculate our limits for two different values of $\rho_\odot$, which also serve the purpose to roughly bracket current uncertainties on $\rho_\odot$, see the discussion in Sec.~\ref{sec:haloDM}.
We give the associated $J$ factors in Table~\ref{tab:Jfact} for $\langle \sigma v\rangle$, and we display their dependence on $\langle \sigma v\rangle$ in Fig.~\ref{fig:Jfact} for a representative case.

\begin{center}
\begin{table}[!h]
	\centering
	\begin{tabular*}{\textwidth}{l @{\extracolsep{\fill}} ccccc}
		\hline
		& & & &\\
		  & \multicolumn{2}{c}{${\rm NFW}$ } & \multicolumn{2}{c}{Core ($\gamma_c\,=\,0.4,\,r_c\,=\,1~{\rm kpc}$)} & \\
       $\rho_\odot~(\rm GeV\, \rm cm^{-3})$ & $0.383$ & 0.55  & 0.383 & 0.55 \\[1.5ex]
		\hline
  & & & &\\
		$J_{\rm Halo}$ ($10^{18}~\rm GeV^2\, \rm cm^{-5}$) & $548$ & $1.13\times\, 10^{3}$ & $8.27$ & $17$ \\[1.5ex]
		$J_{\rm GS, \,\rm NA}$ ($10^{23}$) & $1.15\times\, 10^{5}$ & $2.37\times\, 10^{5}$ & $1.64$ & $3.34$\\[1.5ex]
		$J_{\bigstar \rm heating\,\rm NA}$ ($10^{20}$) & $3.22\times\,10^{5}$ & $6.64\times\, 10^{5}$ & $6.78$ & $14$ \\[1.5ex]
		\hline
	\end{tabular*}
	\caption{$J$ factors for NFW and core halo dark matter profiles, with and without spikes at the Galactic Centre, calculated at two different $\rho_\odot$ values and for $\langle\sigma v\rangle = 0$.}
	\label{tab:Jfact}
\end{table}
\end{center}

\begin{figure}[!h]
    \centering
    \includegraphics[scale =0.6]{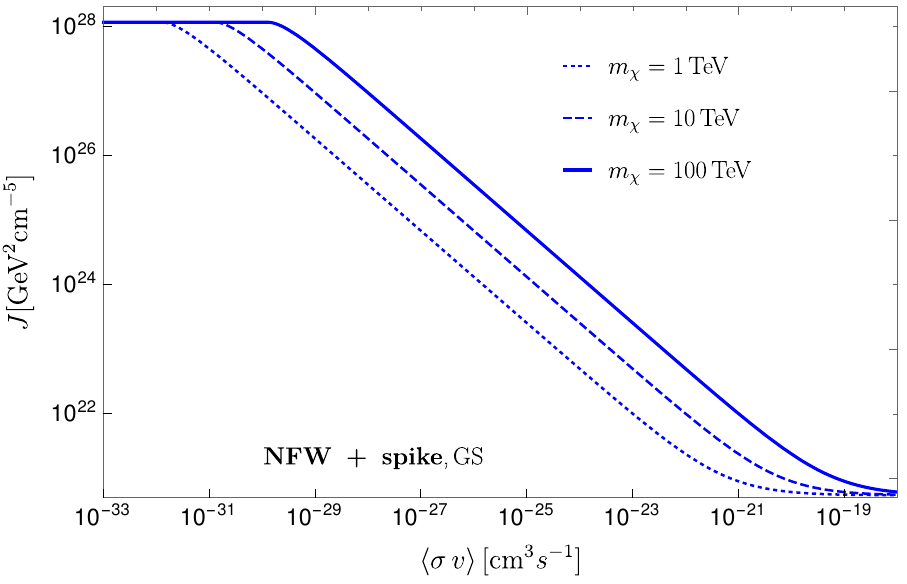}
    \caption{ Variation of the $J$ factor with $\langle\sigma v\rangle$ for local dark matter density $\rho_\odot\,=\,0.383~{\rm GeV\,cm^{-3}}$, $m_\chi\,=\,1,\,10,\,100~{\rm TeV}$ and NFW halo profile with the GS spike benchmark.
  }
    \label{fig:Jfact}
\end{figure}

\section{Results}
In this section, we 
derive limits on the $s$-wave DM annihilation cross section from H.E.S.S. observations of Sgr A*~\cite{HESS:2016pst}, then we discuss their sensitivity to
assumptions about the form of the spike. We begin by discussing how we perform a fit between theoretical predictions and H.E.S.S. data then proceed to show the DM limits we obtain.

\subsection{$\chi^2$-fit to H.E.S.S. data}
We start by deriving bounds on the $\langle\sigma v\rangle$-$m_\chi$ plane under the most conservative approach, i.e. without assuming any astrophysical modeling of the data. We opt for a $\chi^2$-squared fit, which compares the expected signal from DM annihilation to the observed $\gamma$-ray spectrum. By adjusting the parameters of the model, such as the DM mass $m_\chi$ and annihilation cross-section $\langle \sigma v\rangle$, the fit can be used to obtain constraints on the properties of DM. The upper bounds on the annihilation cross-section obtained from the $\chi^2$-squared fit are excluded. We proceed under the assumption that the H.E.S.S. data is 
Gaussian-distributed, we compute the estimator
\begin{equation}
\chi^2 = \displaystyle\sum_{i=1}^N\frac{\left(y_i^{\rm data}-E_{i}^2 \frac{\Phi_i}{\Delta E_{i}}\right)^2}{y_i^{\rm error}}
\end{equation}
where $\Phi_i$, $E_i$ and $\Delta E_i$ are integrated flux, central energy value and width of $i^\textrm{th}$ energy bin. $y_i^{\rm data}$ and $y_i^{\rm error}$ correspond to flux observed and the associated experimental errors of those energy bins for which the flux due to DM annihilation is larger than the integrated flux $\Phi$ in the central $[0.1^\circ,0.1^\circ]$ pixel from GC data of H.E.S.S. from Fig.~3 of Ref.~\cite{HESS:2016pst}, labelled H.E.S.S. J1745-290. 

{Note that the H.E.S.S. collaboration reported only the statistical uncertainties in $y_i^{\rm error}$~\cite{HESS:2016pst}.
However instrument effects of H.E.S.S. (e.g. effective area) could result in systematic uncertainties in the flux measurements \cite{HESS:2006fka}.  Besides uncertainties on the DM density profile, these systematic error in the measurement could affect the limits on DM annihilation.
Taking a nominal value of 20\% systematic flux uncertainty from Ref.~\cite{HESS:2006fka}, we find that the systematic uncertainty exceeds the statistical error for bins with energy less than 10 TeV.  As a result, the effect on DM limits for energies above a few TeV is confined to at most $\sim25\%$; however, for smaller energies, the limits can weaken by a factor of 2 at most. These estimates hold for all the DM annihilation channels, and moving forward we will focus on the simplest case where $y_i^{\rm error}$ is taken as reported in~\cite{HESS:2016pst}.
}

The allowed annihilation cross-section, $\langle\sigma v\rangle$, for a given DM mass at 95\% C.L. corresponds to values of $\chi^2$ smaller than the threshold $\chi^2_{0.05}[N]$, where N is the number of bins in which the model overshoots the data.
This also means that the DM signal in each bin does not exceed the observed central value plus twice the error bar. In this way, we obtain robust, conservative bounds. These bounds could be significantly strengthened with dedicated searches in the future with higher angular resolution such as  that of CTA, as well as by including fits to different astrophysical background  components.

\subsection{H.E.S.S. limits on dark matter annihilation}

The 95\% CL bounds on the  DM annihilation cross-section $\langle\sigma v\rangle$ as a function of DM mass $m_\chi$ in GeV for various DM spike profiles around the GC are shown in Fig.~\ref{fig:HESSlimits} and follow the same colour scheme as the DM profiles shown in Fig.~\ref{fig:spikeprofiles}.
The solid curves are for a  DM local density of $\rho_\odot=0.383\,\textrm{GeV}\textrm{cm}^{-3}$ while the dashed curves are for $\rho_\odot=0.55\,\textrm{GeV}\textrm{cm}^{-3}$ as discussed in Sec.~\ref{sec:haloDM}.
 In the top panels, we show the NFW with DM spike limits  in the bottom panels, we show results for the DM with cored halo profiles.
In the left panels, we show $\chi\overline{\chi}\rightarrow W^{+}W^{-}$ while in the right panels, we show $\chi\overline{\chi}\rightarrow b\overline{b}$.
Note that we focus on these two channels for our main discussion moving forward, however in Appendix~\ref{sec:appendix} we also display our limits on $\langle \sigma v \rangle$ for
the following final states: $\gamma\gamma,\,ZZ$ (see Fig.~\ref{fig:HESSlimitsvectors}), $\mu\overline{\mu},\,\tau\overline{\tau}$ (see Fig.~\ref{fig:leptoniclimits}) and $hh, t\overline{t}$ (see Fig.~\ref{fig:HESSlimitsHiggsTop}).

\medskip

For NFW with spike, we note the strongest bounds come from the GS benchmark shown in blue. For both channels the limit is strongest for $m_\chi>1$~TeV, and lies around $\langle\sigma v \rangle\simeq 10^{-30}\,\textrm{cm}^3\textrm{s}^{-1}$. This is unsurprising given the smaller fluxes observed at higher energies by H.E.S.S.. The next strongest limit comes from the stellar heating  benchmark shown in purple.
Here for the $b\overline{b}$ channel, we see a fairly consistent bound of $\langle\sigma v \rangle\simeq 10^{-24}\,\textrm{cm}^3\textrm{s}^{-1}$ across the whole mass range, however for the $W^{+}W^{-}$ channel, we see the strongest limit of $\langle\sigma v \rangle\simeq 10^{-26}\,\textrm{cm}^3\textrm{s}^{-1}$ occurring for $10^4 \lesssim m_\chi/\textrm{GeV} \lesssim 10^5$. Finally we show the less stellar heating benchmark in dark cyan and the BM benchmark in gray.
The cross-section limits show little mass variation in both the $b\overline{b}$ and $W^+ W^-$ channels, remaining approximately in the range $\langle \sigma v\rangle \simeq 10^{-24}$-$10^{-23}\,\textrm{cm}^3 \textrm{s}^{-1}$. 
{ We find a dramatic impact of the choice of spike parameters on the resulting indirect detection signals from an NFW halo, as for example found also by~\cite{Liu:2022air} for signals of DM much lighter than ours at DAMPE.}

We also examined the effect of increasing the local DM density
from  $\rho_\odot =  0.383\,\textrm{GeV}\,\textrm{cm}^{-3}$ to $\rho_\odot =  0.55\,\textrm{GeV}\,\textrm{cm}^{-3}$,  where the latter value is favoured by local determinations, see Sec.~\ref{sec:haloDM}. Naively this would result in a strengthening of the limits by a factor of $(0.55/0.383)^2 \simeq 2$..
However, this resulted in a  significant improvement in the limits for the GS and stellar heating benchmarks, with the limit improving by almost an order of magnitude for some values of $m_\chi$. This is because the $J$ factor itself increases with a decrease in $\langle\sigma v\rangle$, as discussed in Section~\ref{sec:flux}. { Note that }the bounds are improved by an expected factor of two for the other two benchmarks, where the DM spike density is comparable to DM halo density because annihilations effectively destroy the spike. This suggests that the effect of the DM spike on the limits depends on the benchmark and the DM properties, and should be taken into account when interpreting the results of DM searches.

\begin{figure}[!h]
    \centering
    \includegraphics[width=7cm]{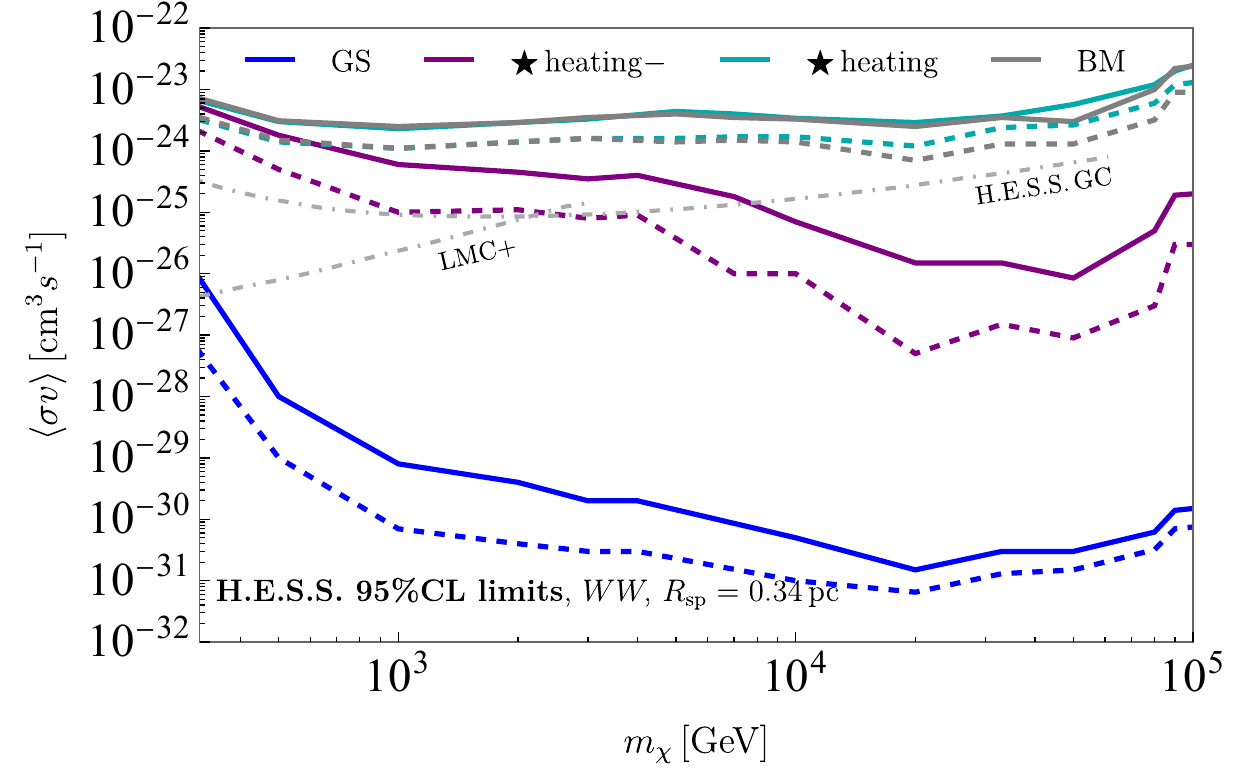}
          \includegraphics[width=7cm]{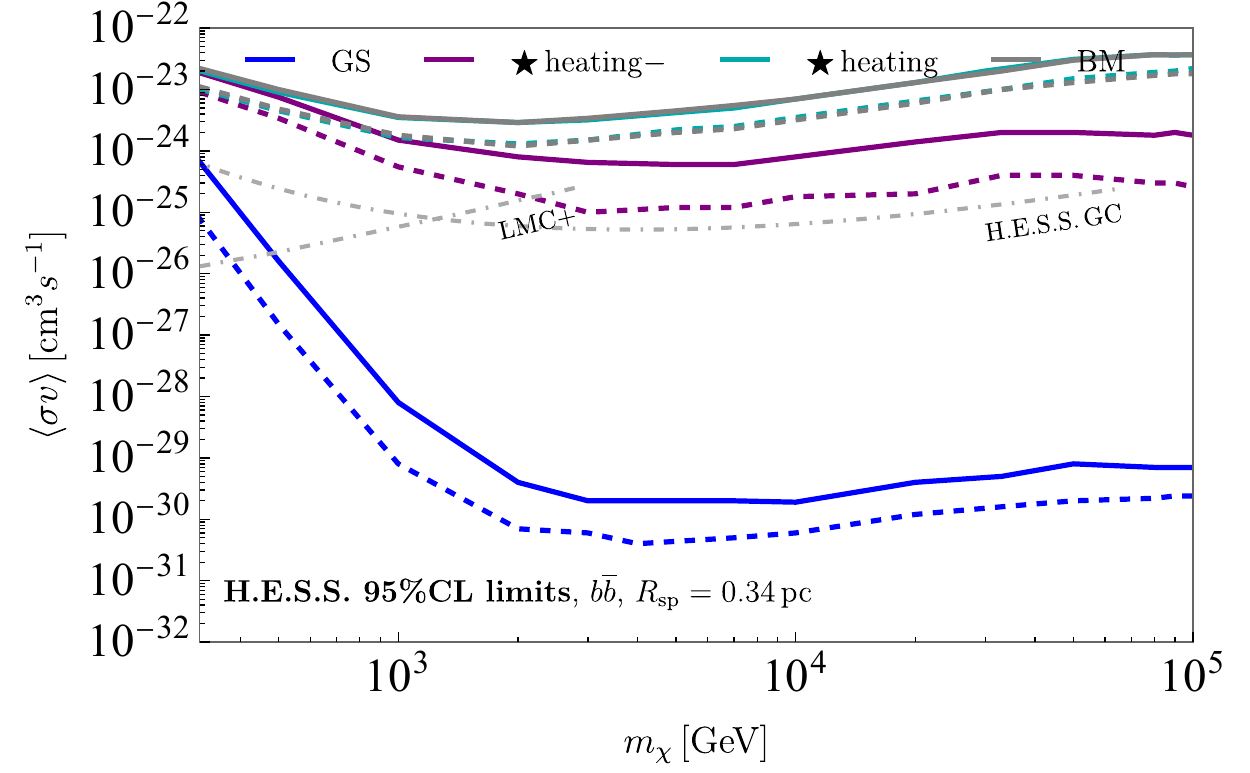}\\
    \includegraphics[width=7cm]{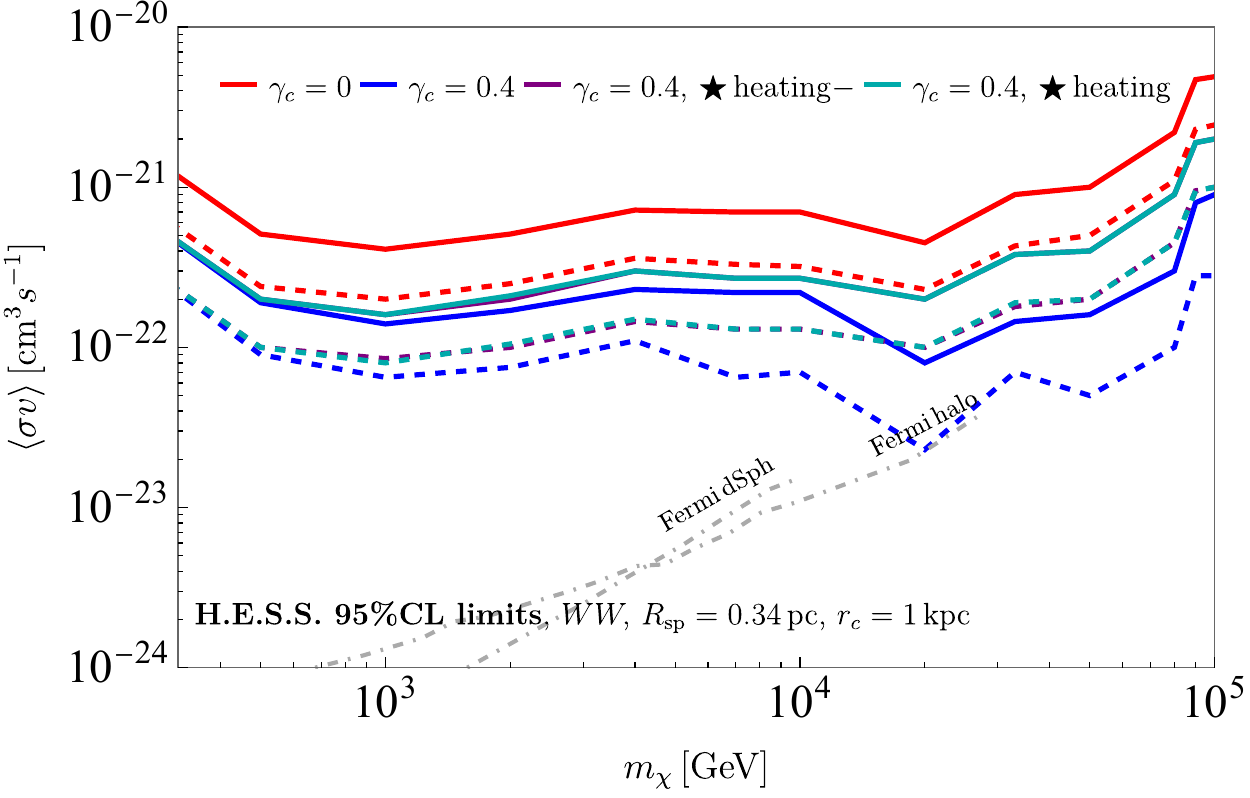}
    \includegraphics[width=7cm]{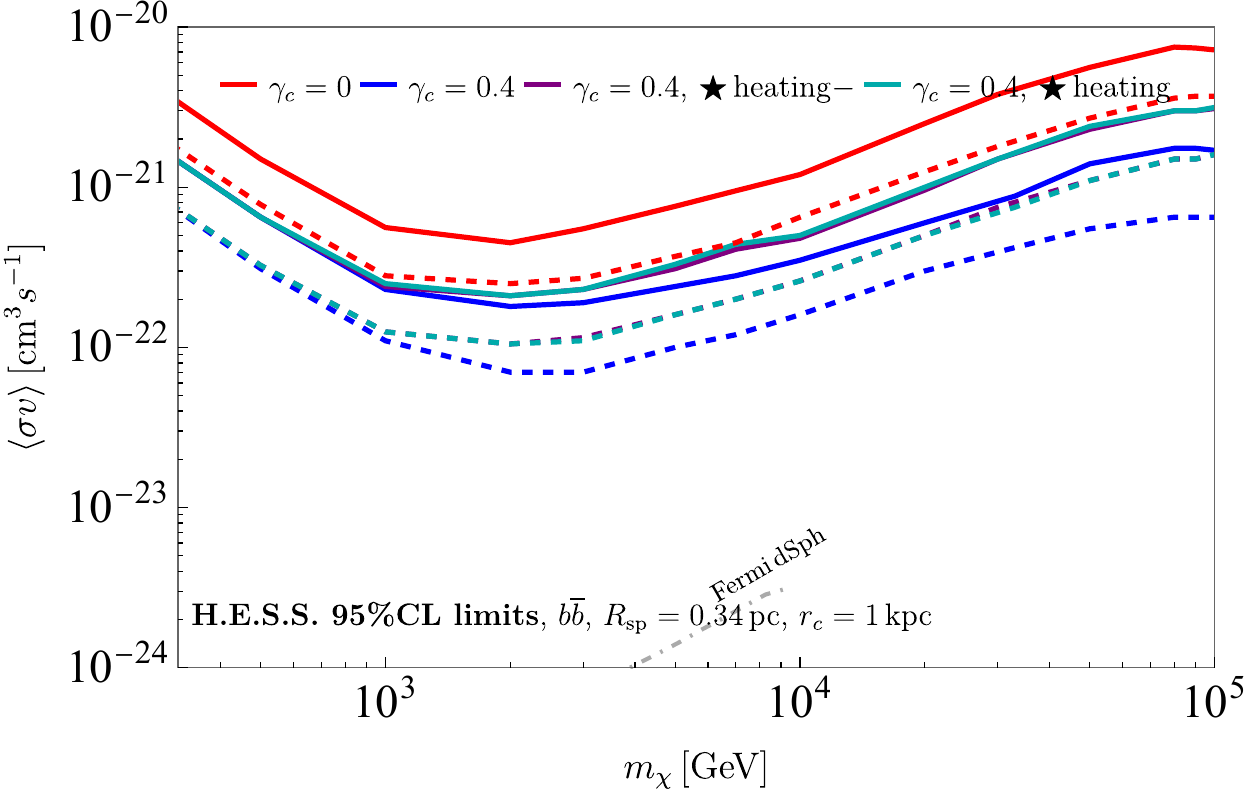}
    \caption{Our results for the 95\% CL limits on the thermally averaged dark matter cross section $\langle \sigma v \rangle$ vs the dark matter mass $m_\chi$, using H.E.S.S. data~\cite{HESS:2016pst}.
    Results for annihilation into $W^{+}W^{-}$ (left) and $b\overline{b}$ (right) and subsequently $\gamma$-rays are shown.
    The continuous lines correspond to the profiles shown in  Fig.~\ref{fig:spikeprofiles}, and we display the NFW case in the top panels and the cored one in the bottom panels.
    The dashed curves are for a local dark matter density $\rho_\odot=0.55\,\textrm{GeV}{\textrm{cm}^{-3}}$ while the solid curves are for $\rho_\odot=0.383\,\textrm{GeV}{\textrm{cm}^{-3}}$, with all other profile parameters kept fixed.
    In each case $R_\textrm{sp}=0.2R_h=0.34$ pc and for cored cases we set a core radius of $r_c=1$ kpc.
    We also show complimentary limits from the Large Magellanic Cloud (LMC+) \cite{Regis:2021glv}, H.E.S.S. Galactic Centre~\cite{HESS:2022ygk} (top panels) and Fermi dSph \cite{Fermi-LAT:2015att} and Fermi halo measurements \cite{Cirelli:2015bda}  (bottom panels) in dot-dashed gray. 
  }
    \label{fig:HESSlimits}
\end{figure}

We also show the complimentary limits for the $W^+ W^-$ channel from the Large Magellanic Cloud (LMC+, where the $+$ means we use the lower side of the band in Fig.~4 of \cite{Regis:2021glv}) and from Fig.~2 of the H.E.S.S. GC flux measurements from Ref.~\cite{HESS:2022ygk}. The LMC+ limit is from a radio search for WIMP DM in the LMC. In this analysis a deep image of the LMC obtained from observations of the Australian Square Kilometre Array Pathfinder (ASKAP) were used as part of the Evolutionary Map of the Universe (EMU) survey. LMC+ is stronger than the NFW and spike profiles we use up to a mass of $m_\chi\simeq3$ TeV bounds except for the GS case, which is still significantly stronger. 
The H.E.S.S. GC results are from a search for DM particle annihilation signals using
new observations from an $\gamma$-ray survey of the Inner Galaxy Survey, at very high energies ($>100$ GeV) performed with the H.E.S.S. array.
The GS  benchmark provides much stronger bounds that H.E.S.S. GC limits across the entire mass range. The limit using the less stellar heating benchmark  is stronger than H.E.S.S. GC  for $m_\chi \gtrsim 8$ TeV.
The stellar heating and BM benchmarks are weaker than the H.E.S.S. GC across the whole mass range.

\medskip

     For the core with spike, we note the strongest bounds come from the core slope $\gamma_c=0.4$ case shown in blue. For both channels the limit is strongest for the $b\overline{b}$ channel the strongest bound on the cross section is $\langle\sigma v \rangle\simeq 10^{-22}\,\textrm{cm}^3\textrm{s}^{-1}$ at around $m_\chi\simeq 1$~TeV. The next strongest limit comes from the $\gamma_c=0.4$ sloped profile with less stellar heating  shown in purple.
     This limit is very close to the $\gamma_c=0.4$ with  stellar heating (shown in dark cyan) for both channels. The weakest bound is the flat core with $\gamma_c=0$ shown in red. For both channels and all profiles, we see that the cross section limits lie between $\simeq10^{-22}$-$10^{-20}\,\textrm{cm}^3\textrm{s}^{-1}$ across the whole mass range. There is far less spread between the strength of the limits in the core with spike relative to the NFW with spike. In contrast to the NFW case, an improvement of a factor of two in bounds for all benchmarks of the core case is observed for larger $\rho_\odot$ (depicted by dashed curves). This is because annihilations effectively destroy the DM spike, resulting in its density becoming comparable to that of the DM halo. 
     
     We show complimentary limits for the $W^+W^-$ channel in the  bottom left panel of Fig.~\ref{fig:HESSlimits} from dwarf spheroidal satellite galaxies (Fermi dSph), from Fig.~8 of Ref.~\cite{Fermi-LAT:2015att} and from diffuse Fermi data (Fermi halo), bottom right panel of Fig.~4 in Ref.~\cite{Cirelli:2015bda}. The Fermi dSph limit is derived from searching for DM annihilation with gamma-ray observations of MW dwarf spheroidal galaxies based on 6 years of Fermi Large Area Telescope data processed with the new Pass 8 event-level analysis. The Fermi halo limit is based on a Burkert DM profile and is not very sensitive to the halo choice.
     Both the Fermi dSph and Fermi halo limits are stronger than the core with spike limits up to masses of $\simeq 10$ TeV and $\simeq 27$ TeV respectively.
     Note that we do not compare the LMC limits with the spike with core case since they are too strong in the low mass range to be illuminating and do not add any additional information when compared with Fermi dSph and Fermi halo bounds. 
     For $m_\chi \gtrsim 20$~TeV and to the best of our knowledge, our core with spike limits are the strongest existing ones.

    \medskip
    
     Considering other limits not displayed in our Figures, 
     $\gamma$-ray observations of dwarf spheroidal galaxies with VERITAS shown in Ref.~\cite{Acharyya:2023ptu} span cross sections of $10^{-24}$-$10^{-22}\,\textrm{cm}^3\textrm{s}^{-1}$ for $b\overline{b}$ and $W^+W^-$ for $m_\chi$ between $1$-$100$ TeV. This is comparable to our NFW with spike BM and additional stellar heating profiles. Observations of the GC with the ANTARES neutrino telescope \cite{ANTARES:2022aoa} sets even weaker cross section limits in the secluded DM scenario. For $m_\chi\simeq 1$ PeV, the ICECUBE experiment \cite{IceCube:2022clp} used high energy neutrino observations to limit the DM annihilation cross section to be smaller $\simeq 10^{-22}\,\textrm{cm}^3\rm {s}^{-1}$
    for the $b\overline{b}$ channel. Using a sample of 31 dwarf irregular (dIrr) galaxies within the field of view of the HAWC Observatory \cite{Alfaro:2023kzf}, the combined annihilation bound is between $10^{-21}$-$10^{-20}\,\textrm{cm}^3\textrm{s}^{-1}$ over DM mass range of 1-100 TeV for both $b\overline{b}$ and $W^+ W^-$ channels. All the NFW with spike and core with spike profiles considered in this work provide stronger limits except for the flat cored profile.


{\subsubsection{Effect of energy resolution}}

\begin{figure}[!h]
    \centering
    \includegraphics[width=7cm]{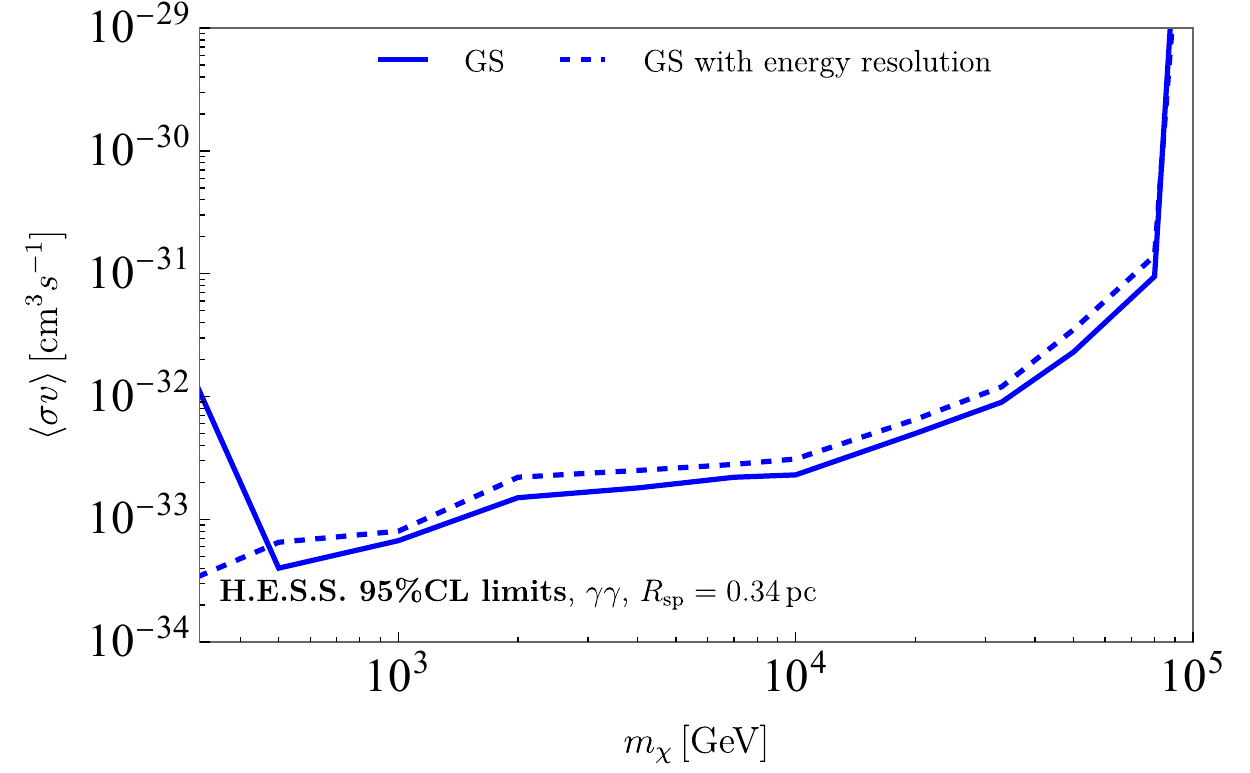}
          \includegraphics[width=7cm]{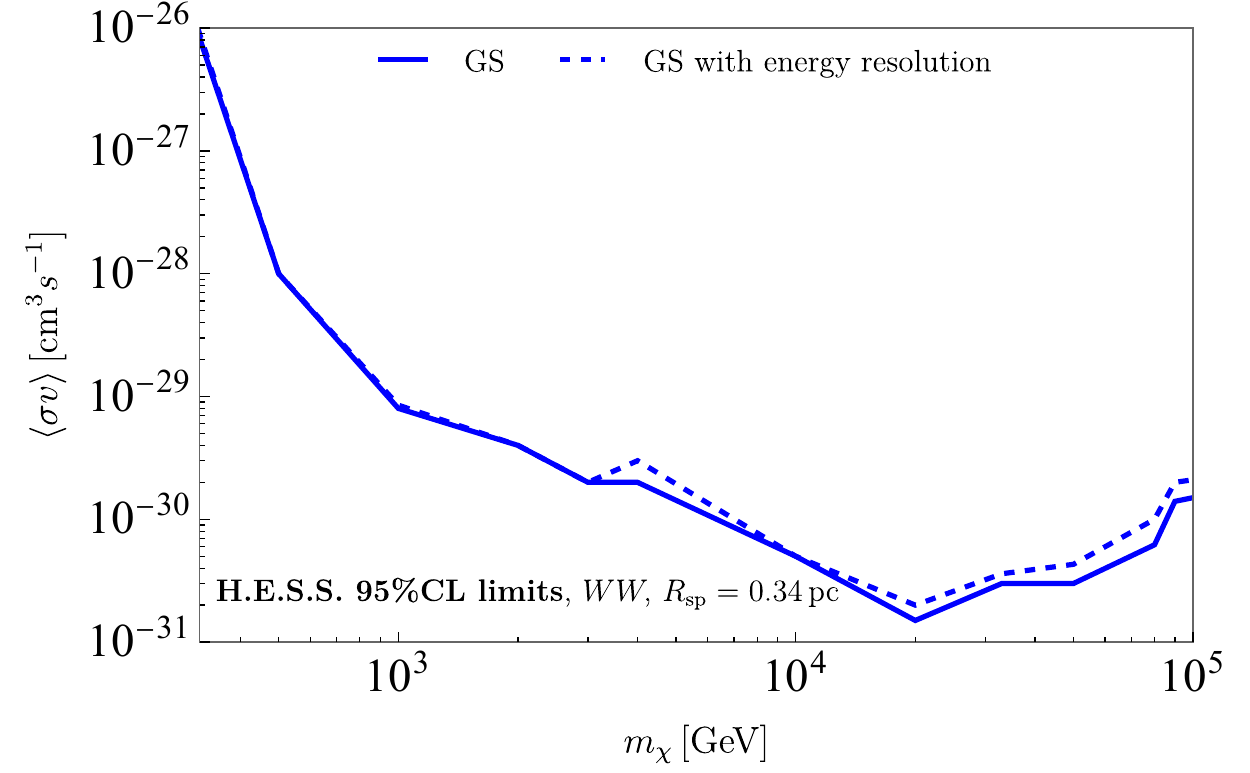}\\
    \caption{{Our results for the 95\% CL limits on the thermally averaged dark matter cross section $\langle \sigma v \rangle$ vs the dark matter mass $m_\chi$, using H.E.S.S. data~\cite{HESS:2016pst}.
    Results for annihilation into $\gamma\gamma$ (left) and $W^{+}W^{-}$ (right) and subsequently $\gamma$-rays are shown.
    The continuous lines correspond to the GS profiles shown in  Fig.~\ref{fig:spikeprofiles}. The dashed lines correspond to the limits obtained using the same GS profile but with a Gaussian energy resolution function of width $15\%$.} 
  }
    \label{fig:energyresolution}
\end{figure}
{We now consider the effect of the energy resolution of a detector as this affects the sensitivity and discrimination of the signal from the background. The energy resolution is the ability of a detector to measure the energy of incoming particles or photons with precision. Here, we express the ratio of the energy uncertainty (or width) to the mean energy of a signal using a Gaussian resolution function. We use the approach highlighted in Section 4.1 of Ref.~\cite{Lefranc:2016fgn} and consider an energy resolution of $\Delta E/E=15\%$ for the GS profile shown in Fig.~\ref{fig:spikeprofiles}. Using this approach for the $\gamma \gamma$ and $W^+W^-$ channels, we get Fig. \ref{fig:energyresolution}, where we show the original limits in solid blue and the limits using the Gaussian resolution function in dashed blue. Observing these limits, we note that including the Gaussian resolution function weakens the limits by at most a factor of a few. Furthermore, we note that the limits are actually strengthened in the lowest DM mass considered for $\gamma\gamma$, because the smeared signal is constrained by lower detector fluxes in the lowest energy bin. We do not consider other channels since the effect of energy resolution will be most pronounced for the channels with the most distinct peaked spectra.}

\section{Conclusion and Outlook}
\label{sec:conclusion}

Recent years have seen the first detections~\cite{HESS:2016pst,Adams:2021kwm,MAGIC:2020kxa} of multi-TeV $\gamma$-rays from the region of the supermassive black hole at the center of the Milky Way galaxy, Sgr A*, and these will be improved and extended in the near future~\cite{LHAASO:2019qtb, CTAConsortium:2019emb,CTA:2020qlo}.
What can one learn from these observations, about DM particle properties and about the DM mass distribution in the Milky Way galaxy?

In this paper, we have performed a step towards addressing this question, by calculating limits on the DM annihilation 
cross-section using H.E.S.S. data from the region of Sgr A*~\cite{HESS:2016pst}.
The main novelties of our work consist in having used these data for the first time, and in having systematically studied the impact of uncertainties on the DM mass density, both in the MW halo and in the spike around Sgr A*.

\medskip

We have given an overview of current knowledge of the DM mass density both in the Milky Way halo, and in the spike around Sgr A*, in Sec.~\ref{sec:DMspikes}. We have defined several benchmarks accordingly, displayed in Fig.~\ref{fig:spikeprofiles}, which can be useful beyond our study. 
Due to the spike softening from DM annihilations, the astrophysical $J$ factors can be very sensitive to $\langle\sigma v\rangle$, unlike the case of most indirect detection searches(see Fig.~\ref{fig:Jfact} for a representative example). 

We have then computed limits on the DM annihilation cross-section for a variety of channels, and displayed them in Figs.~\ref{fig:HESSlimits} ($WW$, $b\bar{b}$), \ref{fig:HESSlimitsvectors} ($\gamma\gamma$, $ZZ$), \ref{fig:leptoniclimits} ($\mu\bar{\mu}$, $\tau\bar{\tau}$) and~\ref{fig:HESSlimitsHiggsTop} ($hh$, $t\bar{t}$). 
To be conservative, we have not assumed any knowledge of the astrophysics responsible for the H.E.S.S. data, i.e. we have compared the DM signal with the observed data assuming it is all due to DM annihilation.
For the case of NFW DM halo profiles, and e.g. for annihilations into $WW$, our new limits are the strongest existing ones for multi-TeV DM not only -as expected- for unperturbed DM spikes, but also if one allows for some softening due to gravitational DM heating by the surrounding stars.

More  generally, our presentation of the results allows us to i) quantitatively grasp the impact of our knowledge of the Milky Way on determining the annihilation properties of DM particles, and to ii) consistently compare the expected signal strengths from the region of Sgr A* with those from other sky regions. 
A future DM signal at some telescope, from any sky region, will then allow us to gain information not only on the DM annihilation properties, but also on its mass density distribution close to Sgr A*.

\medskip

Several interesting avenues of investigation lie ahead. 
While  for definiteness we have considered the case of $s$-wave DM annihilation, it would be straightforward to extend our analysis to velocity-dependent cross-sections.
It would also be very timely to extend our study to DM masses larger than 100~TeV, as annihilating DM in this mass range is attracting increasing attention both from the cosmological and theoretical points of view, see e.g.~\cite{Berlin:2016vnh,Cirelli:2016rnw,Cirelli:2018iax,Hambye:2018qjv,Baldes:2020kam,Baldes:2021aph,Hambye:2020lvy}, and from the observational one.
In particular, new particle physics tools to reliably predict the signals in that mass range~\cite{Bauer:2020jay} are now available, and future analyses of the Sgr A* region { are expected to greatly benefit from the sensitivities to PeV $\gamma$-rays of LHAASO~\cite{LHAASO:2019qtb}, and from the expected angular resolution better than 0.05$^\circ$ above 1 TeV for CTA~\cite{CTAConsortium:2019emb,CTA:2020qlo}.
A precise quantification of these improvements requires a detailed analysis, because of the strong non-linear dependence of the $J$-factor on the tested annihilation cross section along with other possible detector effects, hence we leave it for future work.
}
Finally, multi-TeV DM is an important part of the physics case for future colliders~\cite{Aime:2022flm,FCC:2018byv}, which is potentially enhanced by any associated indirect detection study.

\section*{Acknowledgements}
F.S. thanks Bradley Kavanagh for useful exchanges on DM spikes {and Haibo-Yu for a useful comment on the manuscript}.
 S.B. and D.S. are supported by funding from the European Union's Horizon 2020 research and innovation programme under grant agreement No.~101002846 (ERC CoG ``CosmoChart''). S.B. is also supported by the Initiative Physique des Infinis (IPI), a research training program of the Idex SUPER at Sorbonne Universite.
 F.S. acknowledges support from the Research grant TAsP (Theoretical Astroparticle Physics) funded by Istituto Nazionale di Fisica Nucleare. J.S thanks Emmanuel Moulin for discussions
on angular resolution.

\appendix
\section{DM annihilation limits into $\gamma\gamma, ZZ$, $\mu\mu,\tau\tau$, $hh, tt$}
\label{sec:appendix}
In this appendix, we show our limits on $\langle \sigma v\rangle$ derived from H.E.S.S. data for DM annihilation into $\gamma$-rays through various other intermediate SM states. In Fig.~\ref{fig:HESSlimitsvectors}, Fig.~\ref{fig:leptoniclimits} and finally Fig.~\ref{fig:HESSlimitsHiggsTop} we show annihilation channels into $(\gamma\gamma, ZZ)$, $(\mu\overline{\mu},\tau\overline{\tau})$ and $(hh,t\overline{t}$) for NFW with spike (top panels) and core with spike (bottom panels) respectively. We also show the various cases in the same colour scheme used in Fig.~\ref{fig:HESSlimits}. 

\begin{figure}[!htp]
    \centering
    \includegraphics[width=7cm]{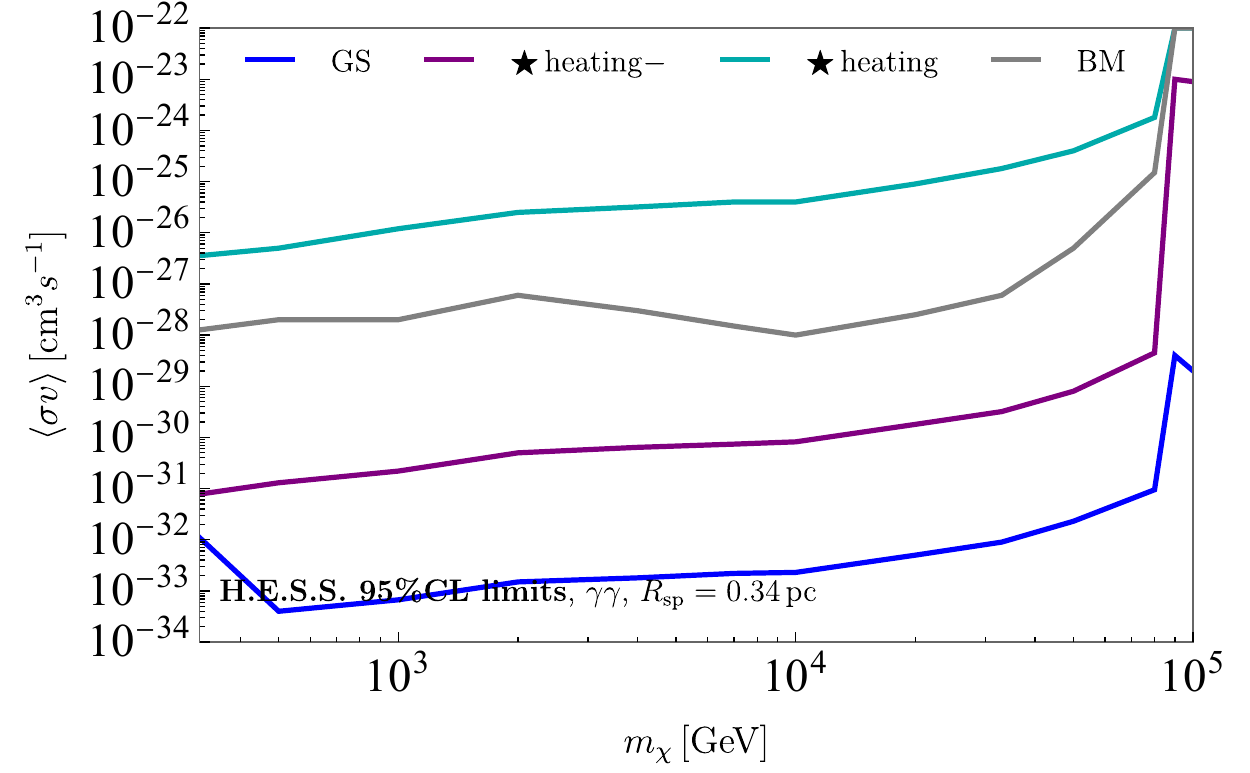}
    \includegraphics[width=7cm]{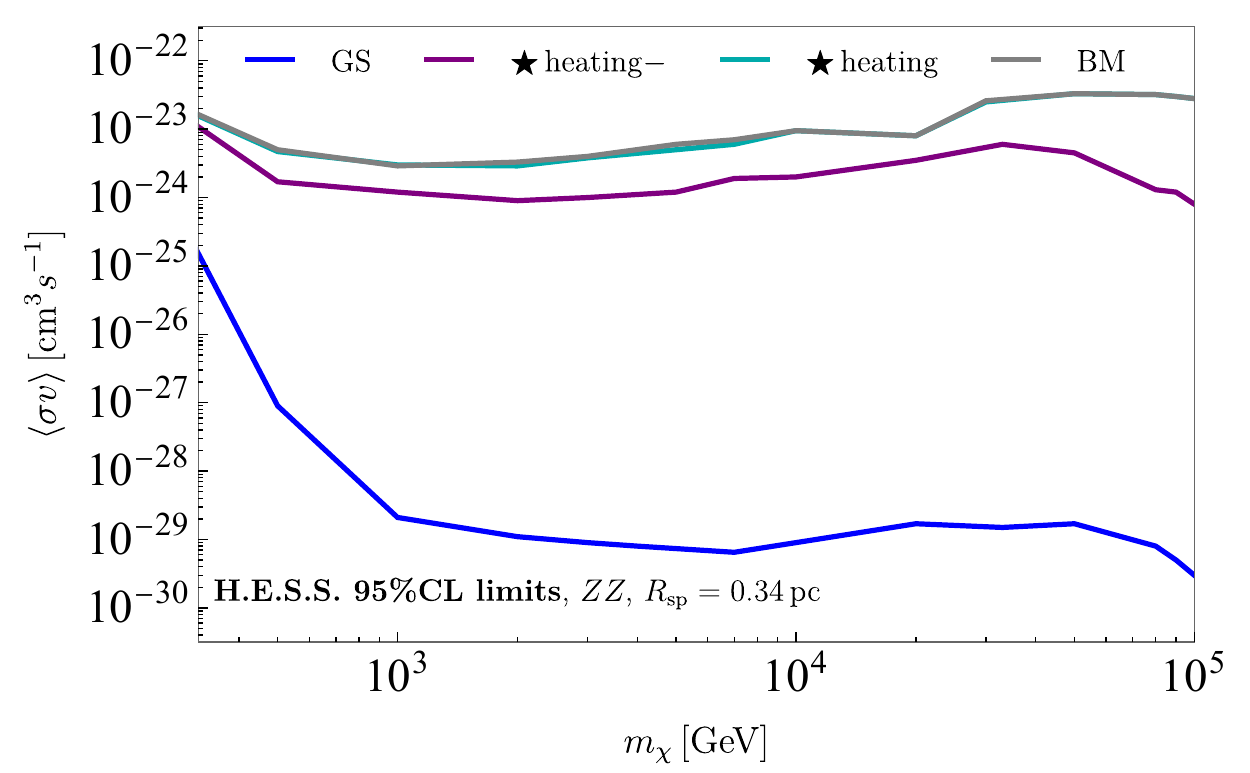}\\
    \includegraphics[width=7cm]{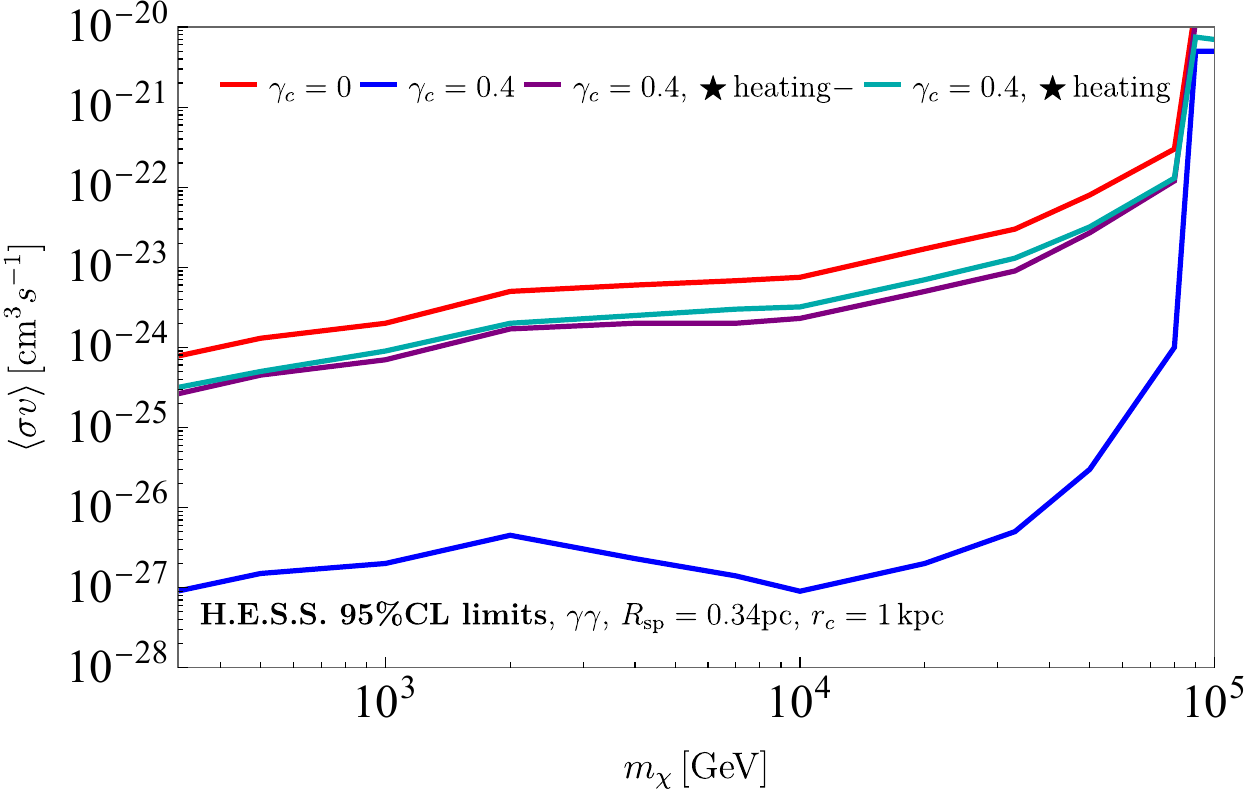}
    \includegraphics[width=7cm]{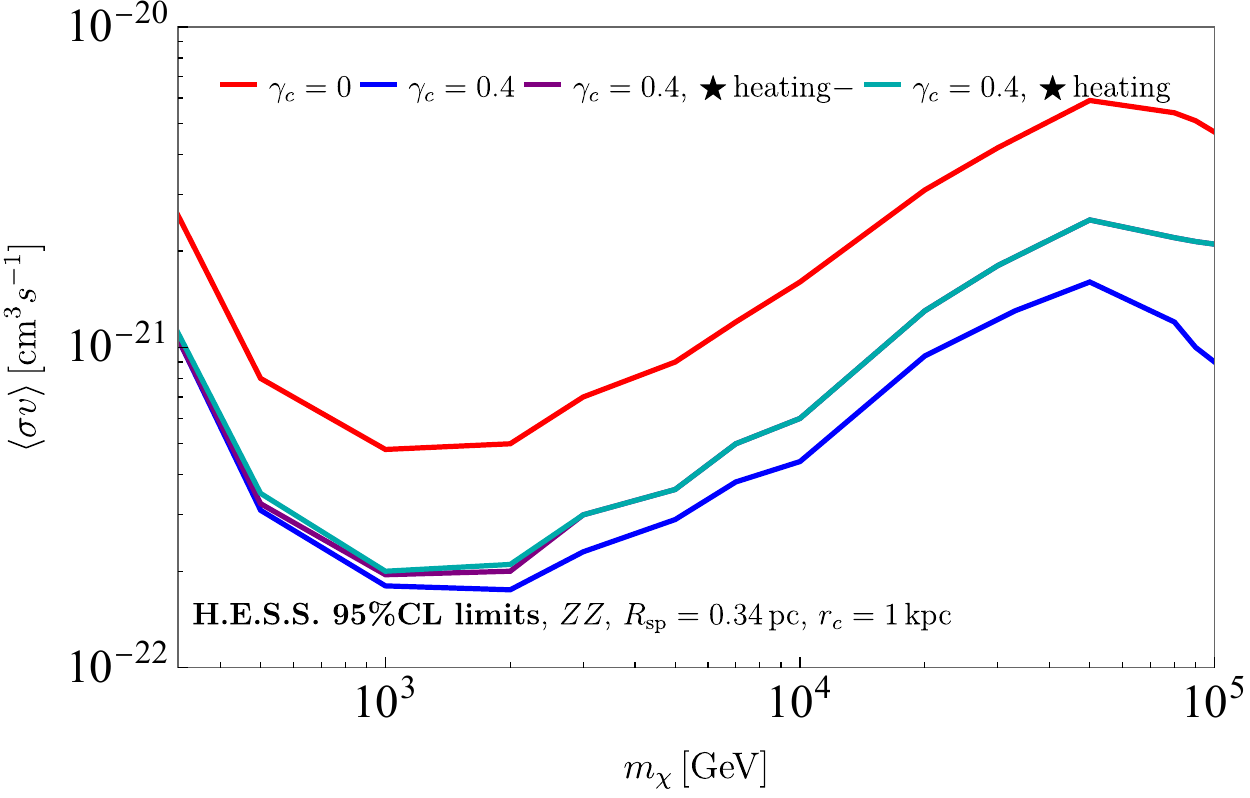} 
    \caption{Our results for the 95\% CL limits on the thermally averaged dark matter cross section $\langle \sigma v \rangle$ vs the dark matter mass $m_\chi$ for DM annihilation into $\gamma\gamma$ (left) and $ZZ$ (right) and for NFW (top) and cored (bottom) DM profiles, using H.E.S.S. data~\cite{HESS:2016pst}. Continuous lines as in Fig.~\ref{fig:HESSlimits}.}
    \label{fig:HESSlimitsvectors}
\end{figure}

\begin{figure}[!htp]
    \centering

     \includegraphics[width=7cm]{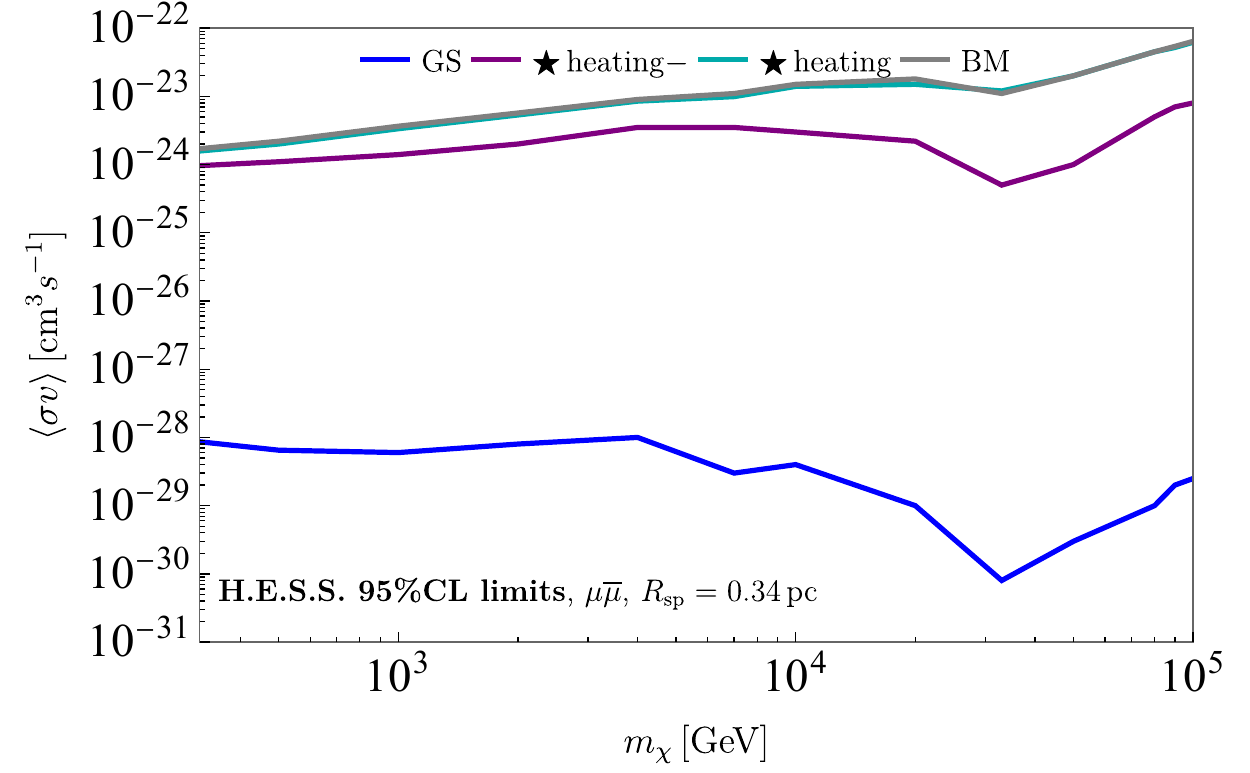}
     \includegraphics[width=7cm]{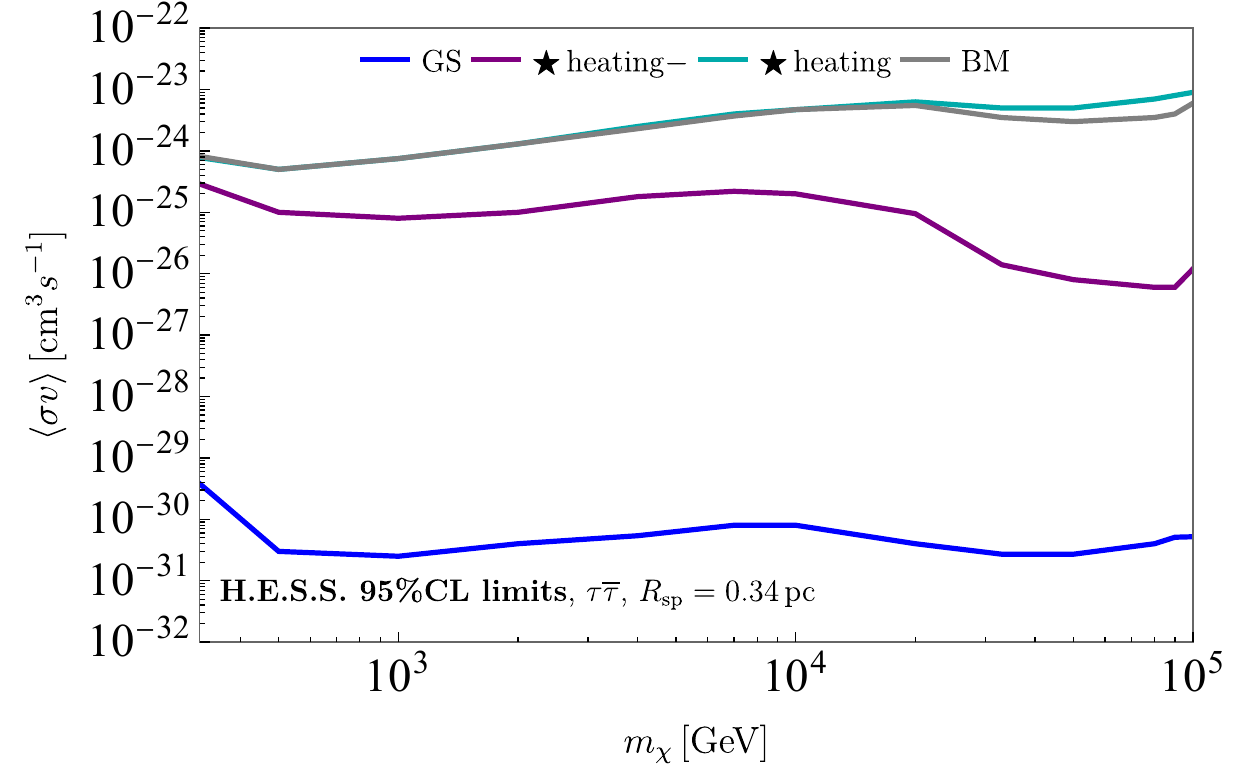}\\
     \includegraphics[width=7cm]{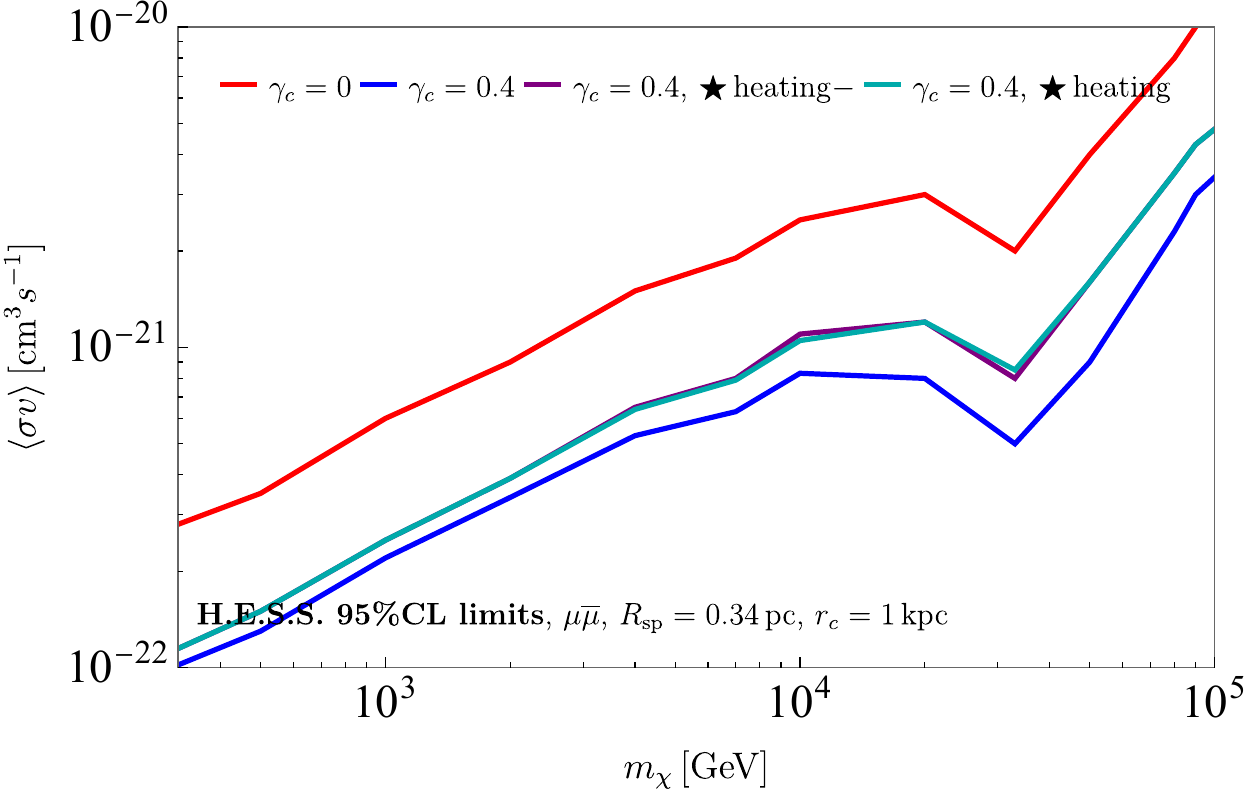}
     \includegraphics[width=7cm]{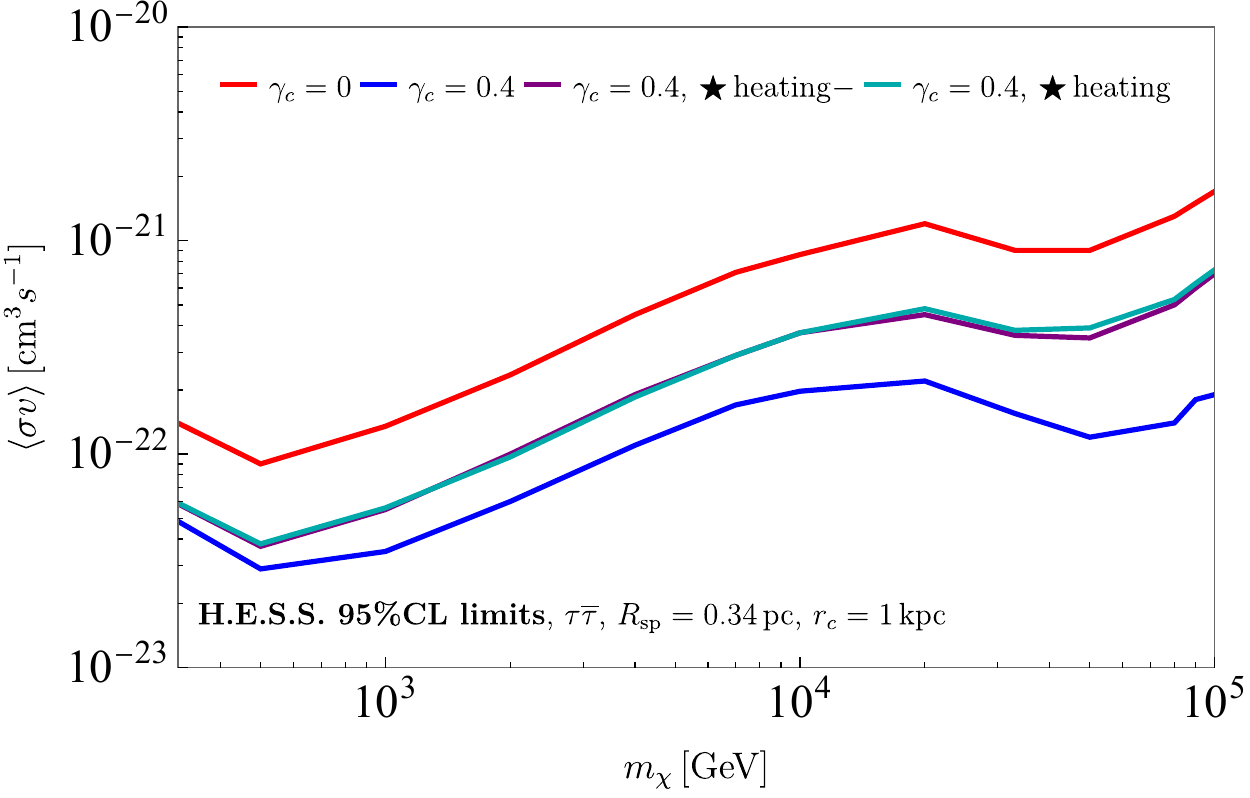}
    \caption{Our results for the 95\% CL limits on the thermally averaged dark matter cross section $\langle \sigma v \rangle$ vs the dark matter mass $m_\chi$ for DM annihilation into $\mu\overline{\mu}$ (left) and $\tau\overline{\tau}$ (right) and for NFW (top) and cored (bottom) DM profiles, using H.E.S.S. data~\cite{HESS:2016pst}.
    Continuous lines as in Fig.~\ref{fig:HESSlimits}.}
    \label{fig:leptoniclimits}
\end{figure}

\begin{figure}[!htp]
    \centering
       \includegraphics[width=7cm]{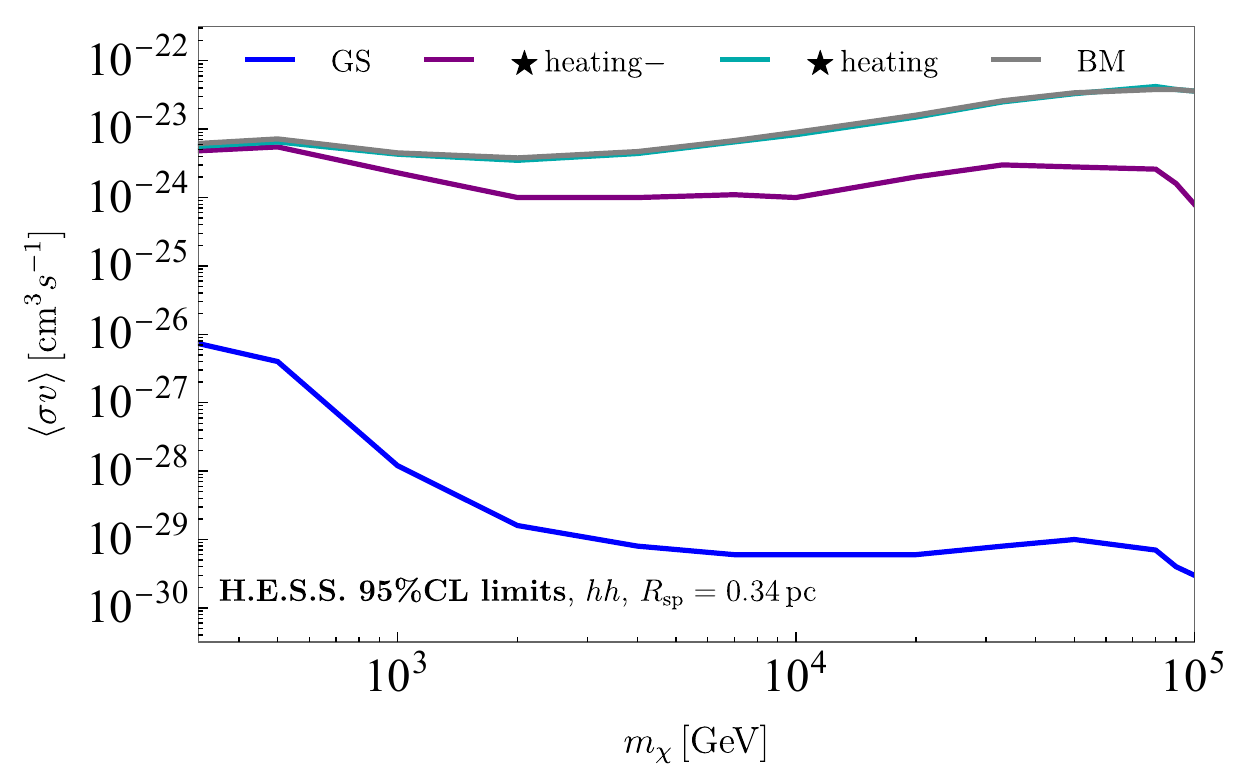}
      \includegraphics[width=7cm]{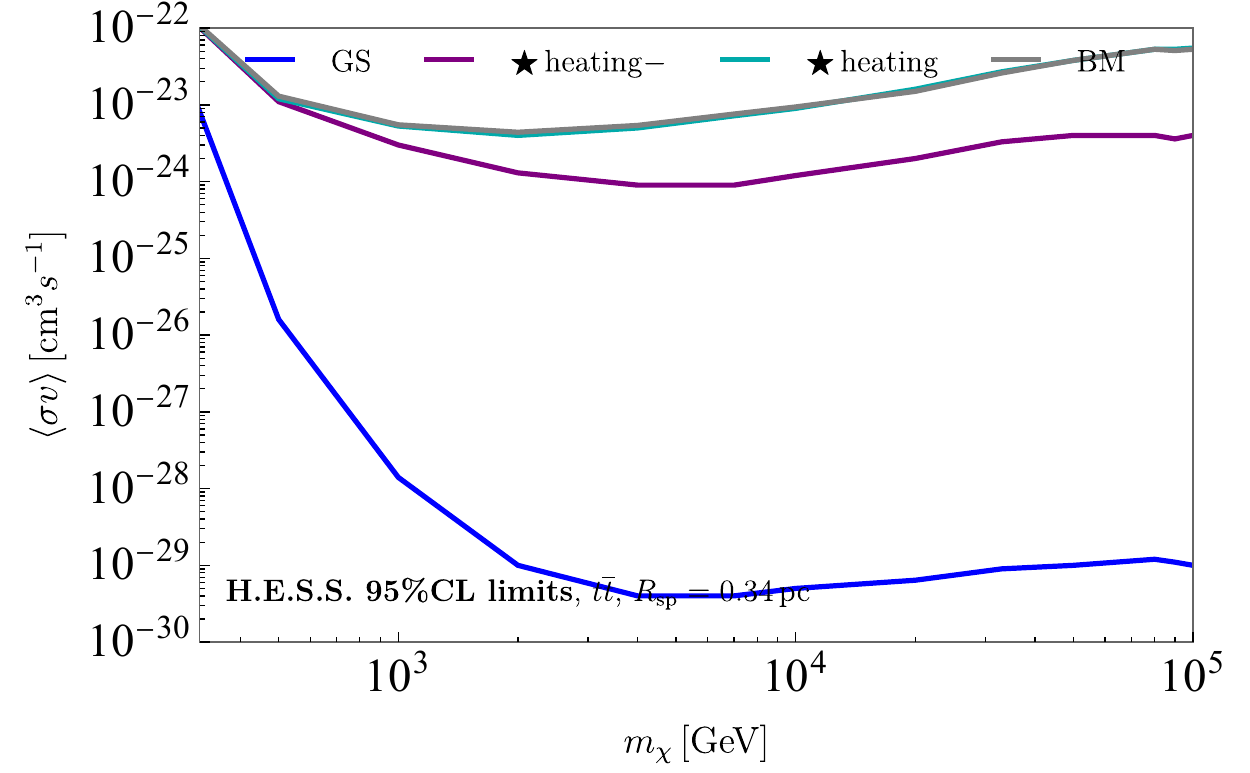}\\
     \includegraphics[width=7cm]{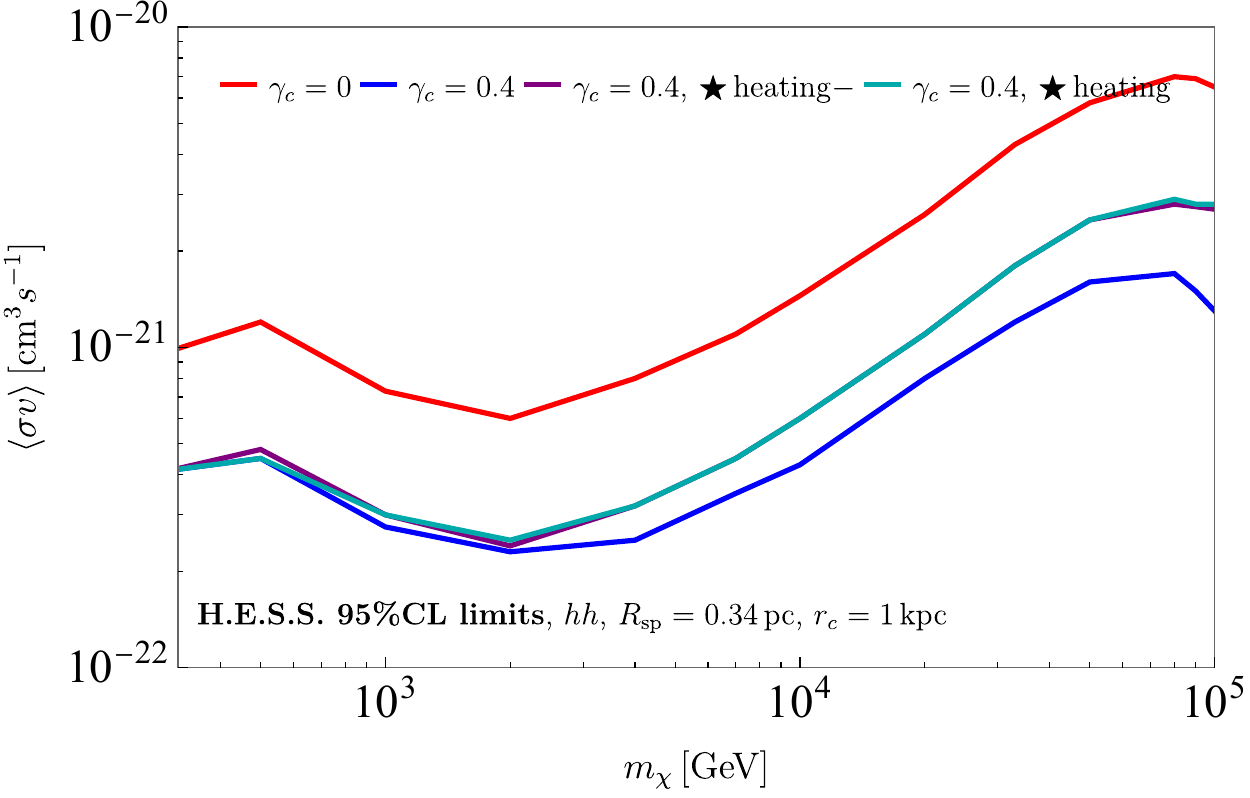}
    \includegraphics[width=7cm]{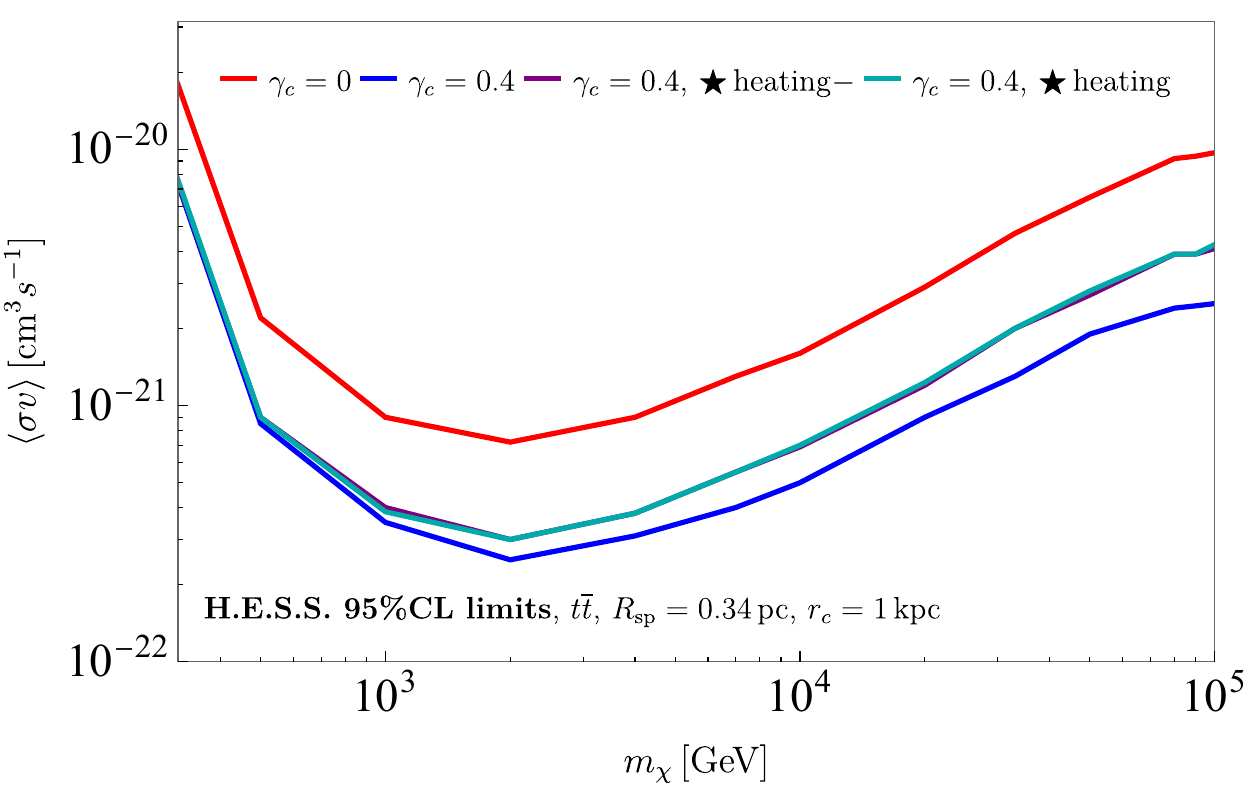}
    \caption{Our results for the 95\% CL limits on the thermally averaged dark matter cross section $\langle \sigma v \rangle$ vs the dark matter mass $m_\chi$ for DM annihilation into $hh$ (left) and $t\overline{t}$ (right) and for NFW (top) and cored (bottom) DM profiles, using H.E.S.S. data~\cite{HESS:2016pst}.
    Continuous lines as in Fig.~\ref{fig:HESSlimits}.}
    \label{fig:HESSlimitsHiggsTop}
\end{figure}

\newpage

\bibliographystyle{JHEP}
\bibliography{refs}

\providecommand{\href}[2]{#2}\begingroup\raggedright\begin{thebibliography}{10}

\bibitem{Boveia:2018yeb}
A.~Boveia and C.~Doglioni, \emph{{Dark Matter Searches at Colliders}},
  \href{http://dx.doi.org/10.1146/annurev-nucl-101917-021008}{\emph{Ann. Rev.
  Nucl. Part. Sci.} {\bf 68} (2018) 429--459},
  [\href{http://arxiv.org/abs/1810.12238}{{\tt 1810.12238}}].

\bibitem{Hinzmann:2022okt}
{\scshape ATLAS, CMS, LHCb} collaboration, A.~Hinzmann, \emph{{Searches for
  Exotica}}, \href{http://dx.doi.org/10.22323/1.398.0035}{\emph{PoS} {\bf
  EPS-HEP2021} (2022) 035}.

\bibitem{Slatyer:2021qgc}
T.~R. Slatyer, \emph{{Les Houches Lectures on Indirect Detection of Dark
  Matter}},
  \href{http://dx.doi.org/10.21468/SciPostPhysLectNotes.53}{\emph{SciPost Phys.
  Lect. Notes} {\bf 53} (2022) 1}, [\href{http://arxiv.org/abs/2109.02696}{{\tt
  2109.02696}}].

\bibitem{Gondolo:1999ef}
P.~Gondolo and J.~Silk, \emph{{Dark matter annihilation at the galactic
  center}}, \href{http://dx.doi.org/10.1103/PhysRevLett.83.1719}{\emph{Phys.
  Rev. Lett.} {\bf 83} (1999) 1719--1722},
  [\href{http://arxiv.org/abs/astro-ph/9906391}{{\tt astro-ph/9906391}}].

\bibitem{HESS:2016pst}
{\scshape H.E.S.S.} collaboration, A.~Abramowski et~al., \emph{{Acceleration of
  petaelectronvolt protons in the Galactic Centre}},
  \href{http://dx.doi.org/10.1038/nature17147}{\emph{Nature} {\bf 531} (2016)
  476}, [\href{http://arxiv.org/abs/1603.07730}{{\tt 1603.07730}}].

\bibitem{Adams:2021kwm}
C.~B. Adams et~al., \emph{{VERITAS Observations of the Galactic Center Region
  at Multi-TeV Gamma-Ray Energies}},
  \href{http://dx.doi.org/10.3847/1538-4357/abf926}{\emph{Astrophys. J.} {\bf
  913} (2021) 115}, [\href{http://arxiv.org/abs/2104.12735}{{\tt 2104.12735}}].

\bibitem{MAGIC:2020kxa}
{\scshape MAGIC, MAGIC Finnish Consortium: Finnish Centre of Astronomy with
  ESO} collaboration, V.~A. Acciari et~al., \emph{{MAGIC observations of the
  diffuse \ensuremath{\gamma}-ray emission in the vicinity of the Galactic
  center}}, \href{http://dx.doi.org/10.1051/0004-6361/201936896}{\emph{Astron.
  Astrophys.} {\bf 642} (2020) A190},
  [\href{http://arxiv.org/abs/2006.00623}{{\tt 2006.00623}}].

\bibitem{LHAASO:2019qtb}
{\scshape LHAASO} collaboration, A.~Addazi et~al., \emph{{The Large High
  Altitude Air Shower Observatory (LHAASO) Science Book (2021 Edition)}},
  {\emph{Chin. Phys. C} {\bf 46} (2022) 035001--035007},
  [\href{http://arxiv.org/abs/1905.02773}{{\tt 1905.02773}}].

\bibitem{CTAConsortium:2019emb}
{\scshape CTA Consortium} collaboration, A.~Viana et~al., \emph{{The Cherenkov
  Telescope Array view of the Galactic Center region}},
  \href{http://dx.doi.org/10.22323/1.358.0817}{\emph{PoS} {\bf ICRC2019} (2021)
  817}, [\href{http://arxiv.org/abs/1908.06162}{{\tt 1908.06162}}].

\bibitem{CTA:2020qlo}
{\scshape CTA} collaboration, A.~Acharyya et~al., \emph{{Sensitivity of the
  Cherenkov Telescope Array to a dark matter signal from the Galactic centre}},
  \href{http://dx.doi.org/10.1088/1475-7516/2021/01/057}{\emph{JCAP} {\bf 01}
  (2021) 057}, [\href{http://arxiv.org/abs/2007.16129}{{\tt 2007.16129}}].

\bibitem{Pato:2015dua}
M.~Pato, F.~Iocco and G.~Bertone, \emph{{Dynamical constraints on the dark
  matter distribution in the Milky Way}},
  \href{http://dx.doi.org/10.1088/1475-7516/2015/12/001}{\emph{JCAP} {\bf 12}
  (2015) 001}, [\href{http://arxiv.org/abs/1504.06324}{{\tt 1504.06324}}].

\bibitem{2020MNRAS.494.4291C}
M.~{Cautun}, A.~{Ben{\'\i}tez-Llambay}, A.~J. {Deason}, C.~S. {Frenk},
  A.~{Fattahi}, F.~A. {G{\'o}mez} et~al., \emph{{The milky way total mass
  profile as inferred from Gaia DR2}},
  \href{http://dx.doi.org/10.1093/mnras/staa1017}{\emph{Mon. Not. Roy. Astron.
  Soc} {\bf 494} (May, 2020) 4291--4313},
  [\href{http://arxiv.org/abs/1911.04557}{{\tt 1911.04557}}].

\bibitem{DiCintio:2014xia}
A.~Di~Cintio, C.~B. Brook, A.~A. Dutton, A.~V. Macci\`o, G.~S. Stinson and
  A.~Knebe, \emph{{A mass-dependent density profile for dark matter haloes
  including the influence of galaxy formation}},
  \href{http://dx.doi.org/10.1093/mnras/stu729}{\emph{Mon. Not. Roy. Astron.
  Soc.} {\bf 441} (2014) 2986--2995},
  [\href{http://arxiv.org/abs/1404.5959}{{\tt 1404.5959}}].

\bibitem{Sameie:2021ang}
O.~Sameie, M.~Boylan-Kolchin, R.~Sanderson, D.~Vargya, P.~F. Hopkins, A.~Wetzel
  et~al., \emph{{The central densities of Milky Way-mass galaxies in cold and
  self-interacting dark matter models}},
  \href{http://dx.doi.org/10.1093/mnras/stab2173}{\emph{Mon. Not. Roy. Astron.
  Soc.} {\bf 507} (2021) 720--729},
  [\href{http://arxiv.org/abs/2102.12480}{{\tt 2102.12480}}].

\bibitem{GRAVITY:2021xju}
{\scshape GRAVITY} collaboration, R.~Abuter et~al., \emph{{Mass distribution in
  the Galactic Center based on interferometric astrometry of multiple stellar
  orbits}}, \href{http://dx.doi.org/10.1051/0004-6361/202142465}{\emph{Astron.
  Astrophys.} {\bf 657} (2022) L12},
  [\href{http://arxiv.org/abs/2112.07478}{{\tt 2112.07478}}].

\bibitem{Heissel:2021pcw}
G.~Hei\ss{}el, T.~Paumard, G.~Perrin and F.~Vincent, \emph{{The dark mass
  signature in the orbit of S2}},
  \href{http://dx.doi.org/10.1051/0004-6361/202142114}{\emph{Astron.
  Astrophys.} {\bf 660} (2022) A13},
  [\href{http://arxiv.org/abs/2112.07778}{{\tt 2112.07778}}].

\bibitem{2019Habibi}
M.~{Habibi}, S.~{Gillessen}, O.~{Pfuhl}, F.~{Eisenhauer}, P.~M. {Plewa},
  S.~{von Fellenberg} et~al., \emph{{Spectroscopic Detection of a Cusp of
  Late-type Stars around the Central Black Hole in the Milky Way}},
  \href{http://dx.doi.org/10.3847/2041-8213/ab03cf}{\emph{\apjl} {\bf 872}
  (Feb., 2019) L15}, [\href{http://arxiv.org/abs/1902.07219}{{\tt
  1902.07219}}].

\bibitem{pechetti2020luminosity}
R.~Pechetti, A.~Seth, N.~Neumayer, I.~Georgiev, N.~Kacharov and M.~den Brok,
  \emph{Luminosity models and density profiles for nuclear star clusters for a
  nearby volume-limited sample of 29 galaxies},
  \href{http://dx.doi.org/10.3847/1538-4357/abaaa7}{\emph{The Astrophysical
  Journal} {\bf 900} (aug, 2020) 32},
  [\href{http://arxiv.org/abs/1911.09686}{{\tt 1911.09686}}].

\bibitem{2020Gallego}
E.~{Gallego-Cano}, R.~{Sch{\"o}del}, F.~{Nogueras-Lara}, H.~{Dong},
  B.~{Shahzamanian}, T.~K. {Fritz} et~al., \emph{{New constraints on the
  structure of the nuclear stellar cluster of the Milky Way from star counts
  and MIR imaging}},
  \href{http://dx.doi.org/10.1051/0004-6361/201935303}{\emph{\aap} {\bf 634}
  (Feb., 2020) A71}, [\href{http://arxiv.org/abs/2001.08182}{{\tt
  2001.08182}}].

\bibitem{Ullio:2001fb}
P.~Ullio, H.~Zhao and M.~Kamionkowski, \emph{{A Dark matter spike at the
  galactic center?}},
  \href{http://dx.doi.org/10.1103/PhysRevD.64.043504}{\emph{Phys. Rev. D} {\bf
  64} (2001) 043504}, [\href{http://arxiv.org/abs/astro-ph/0101481}{{\tt
  astro-ph/0101481}}].

\bibitem{Gnedin:2003rj}
O.~Y. Gnedin and J.~R. Primack, \emph{{Dark Matter Profile in the Galactic
  Center}}, \href{http://dx.doi.org/10.1103/PhysRevLett.93.061302}{\emph{Phys.
  Rev. Lett.} {\bf 93} (2004) 061302},
  [\href{http://arxiv.org/abs/astro-ph/0308385}{{\tt astro-ph/0308385}}].

\bibitem{Bertone:2005hw}
G.~Bertone and D.~Merritt, \emph{{Time-dependent models for dark matter at the
  Galactic Center}},
  \href{http://dx.doi.org/10.1103/PhysRevD.72.103502}{\emph{Phys. Rev. D} {\bf
  72} (2005) 103502}, [\href{http://arxiv.org/abs/astro-ph/0501555}{{\tt
  astro-ph/0501555}}].

\bibitem{Shapiro:2022prq}
S.~L. Shapiro and D.~C. Heggie, \emph{{Effect of stars on the dark matter spike
  around a black hole: A tale of two treatments}},
  \href{http://dx.doi.org/10.1103/PhysRevD.106.043018}{\emph{Phys. Rev. D} {\bf
  106} (2022) 043018}, [\href{http://arxiv.org/abs/2209.08105}{{\tt
  2209.08105}}].

\bibitem{Navarro:2008kc}
J.~F. Navarro, A.~Ludlow, V.~Springel, J.~Wang, M.~Vogelsberger, S.~D.~M. White
  et~al., \emph{{The Diversity and Similarity of Cold Dark Matter Halos}},
  \href{http://dx.doi.org/10.1111/j.1365-2966.2009.15878.x}{\emph{Mon. Not.
  Roy. Astron. Soc.} {\bf 402} (2010) 21},
  [\href{http://arxiv.org/abs/0810.1522}{{\tt 0810.1522}}].

\bibitem{Diemand:2008in}
J.~Diemand, M.~Kuhlen, P.~Madau, M.~Zemp, B.~Moore, D.~Potter et~al.,
  \emph{{Clumps and streams in the local dark matter distribution}},
  \href{http://dx.doi.org/10.1038/nature07153}{\emph{Nature} {\bf 454} (2008)
  735--738}, [\href{http://arxiv.org/abs/0805.1244}{{\tt 0805.1244}}].

\bibitem{Mashchenko:2007jp}
S.~Mashchenko, J.~Wadsley and H.~M.~P. Couchman, \emph{{Stellar Feedback in
  Dwarf Galaxy Formation}},
  \href{http://dx.doi.org/10.1126/science.1148666}{\emph{Science} {\bf 319}
  (2008) 174}, [\href{http://arxiv.org/abs/0711.4803}{{\tt 0711.4803}}].

\bibitem{Maccio:2011ryn}
A.~V. Maccio', G.~Stinson, C.~B. Brook, J.~Wadsley, H.~M.~P. Couchman, S.~Shen
  et~al., \emph{{Halo expansion in cosmological hydro simulations: towards a
  baryonic solution of the cusp/core problem in massive spirals}},
  \href{http://dx.doi.org/10.1088/2041-8205/744/1/L9}{\emph{Astrophys. J.
  Lett.} {\bf 744} (2012) L9}, [\href{http://arxiv.org/abs/1111.5620}{{\tt
  1111.5620}}].

\bibitem{Governato:2012fa}
F.~Governato, A.~Zolotov, A.~Pontzen, C.~Christensen, S.~H. Oh, A.~M. Brooks
  et~al., \emph{{Cuspy No More: How Outflows Affect the Central Dark Matter and
  Baryon Distribution in Lambda CDM Galaxies}},
  \href{http://dx.doi.org/10.1111/j.1365-2966.2012.20696.x}{\emph{Mon. Not.
  Roy. Astron. Soc.} {\bf 422} (2012) 1231--1240},
  [\href{http://arxiv.org/abs/1202.0554}{{\tt 1202.0554}}].

\bibitem{Gnedin:2004cx}
O.~Y. Gnedin, A.~V. Kravtsov, A.~A. Klypin and D.~Nagai, \emph{{Response of
  dark matter halos to condensation of baryons: Cosmological simulations and
  improved adiabatic contraction model}},
  \href{http://dx.doi.org/10.1086/424914}{\emph{Astrophys. J.} {\bf 616} (2004)
  16--26}, [\href{http://arxiv.org/abs/astro-ph/0406247}{{\tt
  astro-ph/0406247}}].

\bibitem{Gnedin:2011uj}
O.~Y. Gnedin, D.~Ceverino, N.~Y. Gnedin, A.~A. Klypin, A.~V. Kravtsov,
  R.~Levine et~al., \emph{{Halo Contraction Effect in Hydrodynamic Simulations
  of Galaxy Formation}},  \href{http://arxiv.org/abs/1108.5736}{{\tt
  1108.5736}}.

\bibitem{Rocha:2012jg}
M.~Rocha, A.~H.~G. Peter, J.~S. Bullock, M.~Kaplinghat, S.~Garrison-Kimmel,
  J.~Onorbe et~al., \emph{{Cosmological Simulations with Self-Interacting Dark
  Matter I: Constant Density Cores and Substructure}},
  \href{http://dx.doi.org/10.1093/mnras/sts514}{\emph{Mon. Not. Roy. Astron.
  Soc.} {\bf 430} (2013) 81--104}, [\href{http://arxiv.org/abs/1208.3025}{{\tt
  1208.3025}}].

\bibitem{Kaplinghat:2013xca}
M.~Kaplinghat, R.~E. Keeley, T.~Linden and H.-B. Yu, \emph{{Tying Dark Matter
  to Baryons with Self-interactions}},
  \href{http://dx.doi.org/10.1103/PhysRevLett.113.021302}{\emph{Phys. Rev.
  Lett.} {\bf 113} (2014) 021302}, [\href{http://arxiv.org/abs/1311.6524}{{\tt
  1311.6524}}].

\bibitem{Navarro:1995iw}
J.~F. Navarro, C.~S. Frenk and S.~D.~M. White, \emph{{The Structure of cold
  dark matter halos}}, \href{http://dx.doi.org/10.1086/177173}{\emph{Astrophys.
  J.} {\bf 462} (1996) 563--575},
  [\href{http://arxiv.org/abs/astro-ph/9508025}{{\tt astro-ph/9508025}}].

\bibitem{mcmillan2016mass}
P.~J. McMillan, \emph{The mass distribution and gravitational potential of the
  milky way}, {\emph{Monthly Notices of the Royal Astronomical Society} (2016)
  stw2759}.

\bibitem{Chan:2015tna}
T.~K. Chan, D.~Kere\v{s}, J.~O\~norbe, P.~F. Hopkins, A.~L. Muratov, C.~A.
  Faucher-Gigu\`ere et~al., \emph{{The impact of baryonic physics on the
  structure of dark matter haloes: the view from the FIRE cosmological
  simulations}}, \href{http://dx.doi.org/10.1093/mnras/stv2165}{\emph{Mon. Not.
  Roy. Astron. Soc.} {\bf 454} (2015) 2981--3001},
  [\href{http://arxiv.org/abs/1507.02282}{{\tt 1507.02282}}].

\bibitem{Mollitor:2014ara}
P.~Mollitor, E.~Nezri and R.~Teyssier, \emph{{Baryonic and dark matter
  distribution in cosmological simulations of spiral galaxies}},
  \href{http://dx.doi.org/10.1093/mnras/stu2466}{\emph{Mon. Not. Roy. Astron.
  Soc.} {\bf 447} (2015) 1353--1369},
  [\href{http://arxiv.org/abs/1405.4318}{{\tt 1405.4318}}].

\bibitem{Evans:2018bqy}
N.~W. Evans, C.~A.~J. O'Hare and C.~McCabe, \emph{{Refinement of the standard
  halo model for dark matter searches in light of the Gaia Sausage}},
  \href{http://dx.doi.org/10.1103/PhysRevD.99.023012}{\emph{Phys. Rev. D} {\bf
  99} (2019) 023012}, [\href{http://arxiv.org/abs/1810.11468}{{\tt
  1810.11468}}].

\bibitem{Buch:2018qdr}
J.~Buch, S.~C.~J. Leung and J.~Fan, \emph{{Using Gaia DR2 to Constrain Local
  Dark Matter Density and Thin Dark Disk}},
  \href{http://dx.doi.org/10.1088/1475-7516/2019/04/026}{\emph{JCAP} {\bf 04}
  (2019) 026}, [\href{http://arxiv.org/abs/1808.05603}{{\tt 1808.05603}}].

\bibitem{Merritt:2003qc}
D.~Merritt, \emph{{Single and binary black holes and their influence on nuclear
  structure}},  in \emph{{Carnegie Observatories Centennial Symposium. 1.
  Coevolution of Black Holes and Galaxies}}, 1, 2003.
\newblock \href{http://arxiv.org/abs/astro-ph/0301257}{{\tt astro-ph/0301257}}.

\bibitem{2009ApJ...698..198G}
K.~{G{\"u}ltekin}, D.~O. {Richstone}, K.~{Gebhardt}, T.~R. {Lauer},
  S.~{Tremaine}, M.~C. {Aller} et~al., \emph{{The M-{\ensuremath{\sigma}} and
  M-L Relations in Galactic Bulges, and Determinations of Their Intrinsic
  Scatter}}, \href{http://dx.doi.org/10.1088/0004-637X/698/1/198}{\emph{\apj}
  {\bf 698} (June, 2009) 198--221}, [\href{http://arxiv.org/abs/0903.4897}{{\tt
  0903.4897}}].

\bibitem{Nampalliwar:2021tyz}
S.~Nampalliwar, S.~Kumar, K.~Jusufi, Q.~Wu, M.~Jamil and P.~Salucci,
  \emph{{Modeling the Sgr A* Black Hole Immersed in a Dark Matter Spike}},
  \href{http://dx.doi.org/10.3847/1538-4357/ac05cc}{\emph{Astrophys. J.} {\bf
  916} (2021) 116}, [\href{http://arxiv.org/abs/2103.12439}{{\tt 2103.12439}}].

\bibitem{Shen:2023kkm}
Z.-Q. Shen, G.-W. Yuan, C.-Z. Jiang, Y.-L.~S. Tsai, Q.~Yuan and Y.-Z. Fan,
  \emph{{Exploring dark matter spike distribution around the Galactic centre
  with stellar orbits}},  \href{http://arxiv.org/abs/2303.09284}{{\tt
  2303.09284}}.

\bibitem{Merritt:2006mt}
D.~Merritt, S.~Harfst and G.~Bertone, \emph{{Collisionally Regenerated Dark
  Matter Structures in Galactic Nuclei}},
  \href{http://dx.doi.org/10.1103/PhysRevD.75.043517}{\emph{Phys. Rev. D} {\bf
  75} (2007) 043517}, [\href{http://arxiv.org/abs/astro-ph/0610425}{{\tt
  astro-ph/0610425}}].

\bibitem{Ahn:2007ty}
E.-J. Ahn, G.~Bertone and D.~Merritt, \emph{{Impact of Astrophysical Processes
  on the Gamma-Ray Background from Dark Matter Annihilations}},
  \href{http://dx.doi.org/10.1103/PhysRevD.76.023517}{\emph{Phys. Rev. D} {\bf
  76} (2007) 023517}, [\href{http://arxiv.org/abs/astro-ph/0703236}{{\tt
  astro-ph/0703236}}].

\bibitem{Merritt:2003qk}
D.~Merritt, \emph{{Evolution of the dark matter distribution at the galactic
  center}}, \href{http://dx.doi.org/10.1103/PhysRevLett.92.201304}{\emph{Phys.
  Rev. Lett.} {\bf 92} (2004) 201304},
  [\href{http://arxiv.org/abs/astro-ph/0311594}{{\tt astro-ph/0311594}}].

\bibitem{Sandick:2016zeg}
P.~Sandick, K.~Sinha and T.~Yamamoto, \emph{{Black Holes, Dark Matter Spikes,
  and Constraints on Simplified Models with $t$-Channel Mediators}},
  \href{http://dx.doi.org/10.1103/PhysRevD.98.035004}{\emph{Phys. Rev. D} {\bf
  98} (2018) 035004}, [\href{http://arxiv.org/abs/1701.00067}{{\tt
  1701.00067}}].

\bibitem{2015Antonini}
F.~{Antonini}, E.~{Barausse} and J.~{Silk}, \emph{{The Coevolution of Nuclear
  Star Clusters, Massive Black Holes, and Their Host Galaxies}},
  \href{http://dx.doi.org/10.1088/0004-637X/812/1/72}{\emph{\apj} {\bf 812}
  (Oct., 2015) 72}, [\href{http://arxiv.org/abs/1506.02050}{{\tt 1506.02050}}].

\bibitem{Lacroix:2018zmg}
T.~Lacroix, \emph{{Dynamical constraints on a dark matter spike at the Galactic
  Centre from stellar orbits}},
  \href{http://dx.doi.org/10.1051/0004-6361/201832652}{\emph{Astron.
  Astrophys.} {\bf 619} (2018) A46},
  [\href{http://arxiv.org/abs/1801.01308}{{\tt 1801.01308}}].

\bibitem{ciucua2022chasing}
I.~Ciuc{\u{a}}, D.~Kawata, Y.-S. Ting, R.~J. Grand, A.~Miglio, M.~Hayden
  et~al., \emph{Chasing the impact of the gaia-sausage-enceladus merger on the
  formation of the milky way thick disc}, {\emph{arXiv preprint
  arXiv:2211.01006} (2022) }.

\bibitem{Vasiliev:2007vh}
E.~Vasiliev, \emph{{Dark matter annihilation near a black hole: Plateau vs.
  weak cusp}}, \href{http://dx.doi.org/10.1103/PhysRevD.76.103532}{\emph{Phys.
  Rev. D} {\bf 76} (2007) 103532}, [\href{http://arxiv.org/abs/0707.3334}{{\tt
  0707.3334}}].

\bibitem{Shapiro:2016ypb}
S.~L. Shapiro and J.~Shelton, \emph{{Weak annihilation cusp inside the dark
  matter spike about a black hole}},
  \href{http://dx.doi.org/10.1103/PhysRevD.93.123510}{\emph{Phys. Rev. D} {\bf
  93} (2016) 123510}, [\href{http://arxiv.org/abs/1606.01248}{{\tt
  1606.01248}}].

\bibitem{Cirelli:2010xx}
M.~Cirelli, G.~Corcella, A.~Hektor, G.~Hutsi, M.~Kadastik, P.~Panci et~al.,
  \emph{{PPPC 4 DM ID: A Poor Particle Physicist Cookbook for Dark Matter
  Indirect Detection}},
  \href{http://dx.doi.org/10.1088/1475-7516/2012/10/E01}{\emph{JCAP} {\bf 03}
  (2011) 051}, [\href{http://arxiv.org/abs/1012.4515}{{\tt 1012.4515}}].

\bibitem{Ciafaloni:2010ti}
P.~Ciafaloni, D.~Comelli, A.~Riotto, F.~Sala, A.~Strumia and A.~Urbano,
  \emph{{Weak Corrections are Relevant for Dark Matter Indirect Detection}},
  \href{http://dx.doi.org/10.1088/1475-7516/2011/03/019}{\emph{JCAP} {\bf 03}
  (2011) 019}, [\href{http://arxiv.org/abs/1009.0224}{{\tt 1009.0224}}].

\bibitem{HESS:2006fka}
{\scshape H.E.S.S.} collaboration, F.~Aharonian et~al., \emph{{Observations of
  the Crab Nebula with H.E.S.S}},
  \href{http://dx.doi.org/10.1051/0004-6361:20065351}{\emph{Astron. Astrophys.}
  {\bf 457} (2006) 899--915},
  [\href{http://arxiv.org/abs/astro-ph/0607333}{{\tt astro-ph/0607333}}].

\bibitem{Liu:2022air}
T.-C. Liu, J.-G. Cheng, Y.-F. Liang and E.-W. Liang, \emph{{Search for
  gamma-ray line signals around the black hole at the galactic center with
  DAMPE observation}},
  \href{http://dx.doi.org/10.1007/s11433-022-1890-0}{\emph{Sci. China Phys.
  Mech. Astron.} {\bf 65} (2022) 269512},
  [\href{http://arxiv.org/abs/2203.08078}{{\tt 2203.08078}}].

\bibitem{Regis:2021glv}
M.~Regis et~al., \emph{{The EMU view of the Large Magellanic Cloud: troubles
  for sub-TeV WIMPs}},
  \href{http://dx.doi.org/10.1088/1475-7516/2021/11/046}{\emph{JCAP} {\bf 11}
  (2021) 046}, [\href{http://arxiv.org/abs/2106.08025}{{\tt 2106.08025}}].

\bibitem{HESS:2022ygk}
{\scshape H.E.S.S.} collaboration, H.~Abdalla et~al., \emph{{Search for Dark
  Matter Annihilation Signals in the H.E.S.S. Inner Galaxy Survey}},
  \href{http://dx.doi.org/10.1103/PhysRevLett.129.111101}{\emph{Phys. Rev.
  Lett.} {\bf 129} (2022) 111101}, [\href{http://arxiv.org/abs/2207.10471}{{\tt
  2207.10471}}].

\bibitem{Fermi-LAT:2015att}
{\scshape Fermi-LAT} collaboration, M.~Ackermann et~al., \emph{{Searching for
  Dark Matter Annihilation from Milky Way Dwarf Spheroidal Galaxies with Six
  Years of Fermi Large Area Telescope Data}},
  \href{http://dx.doi.org/10.1103/PhysRevLett.115.231301}{\emph{Phys. Rev.
  Lett.} {\bf 115} (2015) 231301}, [\href{http://arxiv.org/abs/1503.02641}{{\tt
  1503.02641}}].

\bibitem{Cirelli:2015bda}
M.~Cirelli, T.~Hambye, P.~Panci, F.~Sala and M.~Taoso, \emph{{Gamma ray tests
  of Minimal Dark Matter}},
  \href{http://dx.doi.org/10.1088/1475-7516/2015/10/026}{\emph{JCAP} {\bf 10}
  (2015) 026}, [\href{http://arxiv.org/abs/1507.05519}{{\tt 1507.05519}}].

\bibitem{Acharyya:2023ptu}
A.~Acharyya et~al., \emph{{Search for Ultraheavy Dark Matter from Observations
  of Dwarf Spheroidal Galaxies with VERITAS}},
  \href{http://arxiv.org/abs/2302.08784}{{\tt 2302.08784}}.

\bibitem{ANTARES:2022aoa}
{\scshape ANTARES} collaboration, A.~Albert et~al., \emph{{Search for secluded
  dark matter towards the Galactic Centre with the ANTARES neutrino
  telescope}},
  \href{http://dx.doi.org/10.1088/1475-7516/2022/06/028}{\emph{JCAP} {\bf 06}
  (2022) 028}, [\href{http://arxiv.org/abs/2203.06029}{{\tt 2203.06029}}].

\bibitem{IceCube:2022clp}
{\scshape IceCube} collaboration, R.~Abbasi et~al., \emph{{Searches for
  Connections between Dark Matter and High-Energy Neutrinos with IceCube}},
  \href{http://arxiv.org/abs/2205.12950}{{\tt 2205.12950}}.

\bibitem{Alfaro:2023kzf}
R.~Alfaro et~al., \emph{{Searching for TeV Dark Matter in Irregular dwarf
  galaxies with HAWC Observatory}},
  \href{http://arxiv.org/abs/2302.07929}{{\tt 2302.07929}}.

\bibitem{Lefranc:2016fgn}
V.~Lefranc, E.~Moulin, P.~Panci, F.~Sala and J.~Silk, \emph{{Dark Matter in
  $\gamma$ lines: Galactic Center vs dwarf galaxies}},
  \href{http://dx.doi.org/10.1088/1475-7516/2016/09/043}{\emph{JCAP} {\bf 09}
  (2016) 043}, [\href{http://arxiv.org/abs/1608.00786}{{\tt 1608.00786}}].

\bibitem{Berlin:2016vnh}
A.~Berlin, D.~Hooper and G.~Krnjaic, \emph{{PeV-Scale Dark Matter as a Thermal
  Relic of a Decoupled Sector}},
  \href{http://dx.doi.org/10.1016/j.physletb.2016.06.037}{\emph{Phys. Lett. B}
  {\bf 760} (2016) 106--111}, [\href{http://arxiv.org/abs/1602.08490}{{\tt
  1602.08490}}].

\bibitem{Cirelli:2016rnw}
M.~Cirelli, P.~Panci, K.~Petraki, F.~Sala and M.~Taoso, \emph{{Dark Matter's
  secret liaisons: phenomenology of a dark U(1) sector with bound states}},
  \href{http://dx.doi.org/10.1088/1475-7516/2017/05/036}{\emph{JCAP} {\bf 05}
  (2017) 036}, [\href{http://arxiv.org/abs/1612.07295}{{\tt 1612.07295}}].

\bibitem{Cirelli:2018iax}
M.~Cirelli, Y.~Gouttenoire, K.~Petraki and F.~Sala, \emph{{Homeopathic Dark
  Matter, or how diluted heavy substances produce high energy cosmic rays}},
  \href{http://dx.doi.org/10.1088/1475-7516/2019/02/014}{\emph{JCAP} {\bf 02}
  (2019) 014}, [\href{http://arxiv.org/abs/1811.03608}{{\tt 1811.03608}}].

\bibitem{Hambye:2018qjv}
T.~Hambye, A.~Strumia and D.~Teresi, \emph{{Super-cool Dark Matter}},
  \href{http://dx.doi.org/10.1007/JHEP08(2018)188}{\emph{JHEP} {\bf 08} (2018)
  188}, [\href{http://arxiv.org/abs/1805.01473}{{\tt 1805.01473}}].

\bibitem{Baldes:2020kam}
I.~Baldes, Y.~Gouttenoire and F.~Sala, \emph{{String Fragmentation in
  Supercooled Confinement and Implications for Dark Matter}},
  \href{http://dx.doi.org/10.1007/JHEP04(2021)278}{\emph{JHEP} {\bf 04} (2021)
  278}, [\href{http://arxiv.org/abs/2007.08440}{{\tt 2007.08440}}].

\bibitem{Baldes:2021aph}
I.~Baldes, Y.~Gouttenoire, F.~Sala and G.~Servant, \emph{{Supercool composite
  Dark Matter beyond 100 TeV}},
  \href{http://dx.doi.org/10.1007/JHEP07(2022)084}{\emph{JHEP} {\bf 07} (2022)
  084}, [\href{http://arxiv.org/abs/2110.13926}{{\tt 2110.13926}}].

\bibitem{Hambye:2020lvy}
T.~Hambye, M.~Lucca and L.~Vanderheyden, \emph{{Dark matter as a heavy thermal
  hot relic}},
  \href{http://dx.doi.org/10.1016/j.physletb.2020.135553}{\emph{Phys. Lett. B}
  {\bf 807} (2020) 135553}, [\href{http://arxiv.org/abs/2003.04936}{{\tt
  2003.04936}}].

\bibitem{Bauer:2020jay}
C.~W. Bauer, N.~L. Rodd and B.~R. Webber, \emph{{Dark matter spectra from the
  electroweak to the Planck scale}},
  \href{http://dx.doi.org/10.1007/JHEP06(2021)121}{\emph{JHEP} {\bf 06} (2021)
  121}, [\href{http://arxiv.org/abs/2007.15001}{{\tt 2007.15001}}].

\bibitem{Aime:2022flm}
C.~Aime et~al., \emph{{Muon Collider Physics Summary}},
  \href{http://arxiv.org/abs/2203.07256}{{\tt 2203.07256}}.

\bibitem{FCC:2018byv}
{\scshape FCC} collaboration, A.~Abada et~al., \emph{{FCC Physics
  Opportunities}: {Future Circular Collider Conceptual Design Report Volume
  1}}, \href{http://dx.doi.org/10.1140/epjc/s10052-019-6904-3}{\emph{Eur. Phys.
  J. C} {\bf 79} (2019) 474}.

\end{thebibliography}\endgroup
\end{document}